\documentclass[twocolumn,english,pra,aps,superscriptaddress,floatfix]{revtex4-1}

\usepackage{amsthm}
\usepackage{amsmath}
\usepackage{graphicx}
\usepackage{amssymb}
\usepackage{bm}
\usepackage{latexsym}

\makeatletter
\@ifundefined{textcolor}{}
{%
 \definecolor{BLACK}{gray}{0}
 \definecolor{WHITE}{gray}{1}
 \definecolor{RED}{rgb}{1,0,0}
 \definecolor{GREEN}{rgb}{0,1,0}
 \definecolor{BLUE}{rgb}{0,0,1}
 \definecolor{CYAN}{cmyk}{1,0,0,0}
 \definecolor{MAGENTA}{cmyk}{0,1,0,0}
 \definecolor{YELLOW}{cmyk}{0,0,1,0}
 }

\makeatother

\usepackage{babel}

\begin{document}

\author{N. Miladinovic}
\affiliation{Department of Physics and Astronomy, McMaster University, 1280 Main
St.\ W., Hamilton, ON, L8S 4M1, Canada} 
\author{F. Hasan}
\affiliation{Department of Physics and Astronomy, McMaster University, 1280 Main
St.\ W., Hamilton, ON, L8S 4M1, Canada} 
\author{N. Chisholm}
\affiliation{School of Engineering and Applied Sciences, Harvard University, Cambridge, Massachusetts, 02138, USA} 
\author{I. E. Linnington}
\affiliation{Department of Physics and Astronomy, McMaster University, 1280 Main
St.\ W., Hamilton, ON, L8S 4M1, Canada} 
\author{E. A. Hinds}
\affiliation{Centre for Cold Matter, Imperial College, Prince Consort Road, London, SW7 2AZ, United Kingdom}\author{D.\ H.\ J.\ O'Dell}
\affiliation{Department of Physics and Astronomy, McMaster University, 1280 Main
St.\ W., Hamilton, ON, L8S 4M1, Canada}

\title{Adiabatic transfer of light in a double cavity and the optical Landau-Zener problem}

\begin{abstract}
We analyze the evolution of an electromagnetic field inside a double cavity when the difference in length between the two cavities is changed, e.g. by translating the common mirror. We find that this allows photons to be moved deterministically from one cavity to the other. We are able to obtain the conditions for adiabatic transfer by first mapping the Maxwell wave equation for the electric field onto a Schr\"{o}dinger-like wave equation, and then using the Landau-Zener result for the transition probability at an avoided crossing. Our analysis reveals that 
this mapping only rigorously holds when the two cavities are weakly coupled (i.e.\ in the regime of a highly reflective common mirror), and that, generally speaking, care is required when attempting a hamiltonian description of cavity electrodynamics with time-dependent boundary conditions.
\end{abstract}


\maketitle

\section{Introduction}
\label{sec:intro}

One of the outstanding challenges facing quantum information proposals based on cavity quantum electrodynamics (CQED) is the development of a method for transferring photons between different cavities \cite{roadmap}. With a robust photon transfer method, a cavity containing one or more atoms can act as a single node in a quantum network made up of many such cavities connected by optical fibres \cite{kimble}. Quantum networks provide a way to solve the problem of scalability in quantum information processing, scalability being one of the Divincenzo criteria for the realization of a quantum computer \cite{divincenzo00}.

A scheme for the efficient transfer of photons between distant atoms has been put forward by Cirac \textit{et al.}\ \cite{cirac97}. They considered what happens when a photon wave packet is emitted by an atom into a high-Q optical cavity which is connected, via an optical fibre, to a second cavity containing a second atom. A generic wave packet reflects from the highly reflective mirror of the second cavity, thus leaving the wave packet to slosh around inside the cavity-fibre system. However, Cirac \textit{et al.}\ realized that if the emitted wave packet was time-symmetric then it could be absorbed by the second atom with high efficiency. The coherent exchange of photons between an atom and a cavity mode can be controlled by time-dependent laser pulses applied through the side of a cavity \cite{boozer07}, and this allows the temporal shape of the wave packet to be controlled.  

A different approach has been suggested by Pellizari  \cite{pellizzari97,lo98}. In analogy with stimulated Raman adiabatic passage (STIRAP) \cite{bergmann98,vitanov01}, which uses two or more control lasers to transfer an atom between two internal states via an intermediate state which is never populated, Pellizari proposed that laser pulses applied to the two atoms could be used to control the transfer of photons between them along the fibre, the role of the intermediate state being played by the cavity and fibre photon states.  The appeal of an adiabatic passage method lies in the fact that the details of the pulse shape and timing are not important. Furthermore, because the cavity and fibre modes are only virtually populated, the scheme is insensitive to cavity and fibre losses.  Inspired by \cite{pellizzari97}, the theory of adiabatic passage of photons between two atoms via cavities and optical fibres has been developed by a number of groups, see e.g., \cite{enk99,serafini06,song07}.  We also note that the basic concept of STIRAP has also been discussed in the context of the adiabatic passage of matter waves, such as atoms and electrons, where it is referred to as coherent tunneling adiabatic passage (CTAP). For example, references  \cite{eckert04} and \cite{rab08} consider the adiabatic transfer of ultracold atoms along a chain of traps. 

\begin{figure}
\includegraphics[width=1.0\columnwidth]{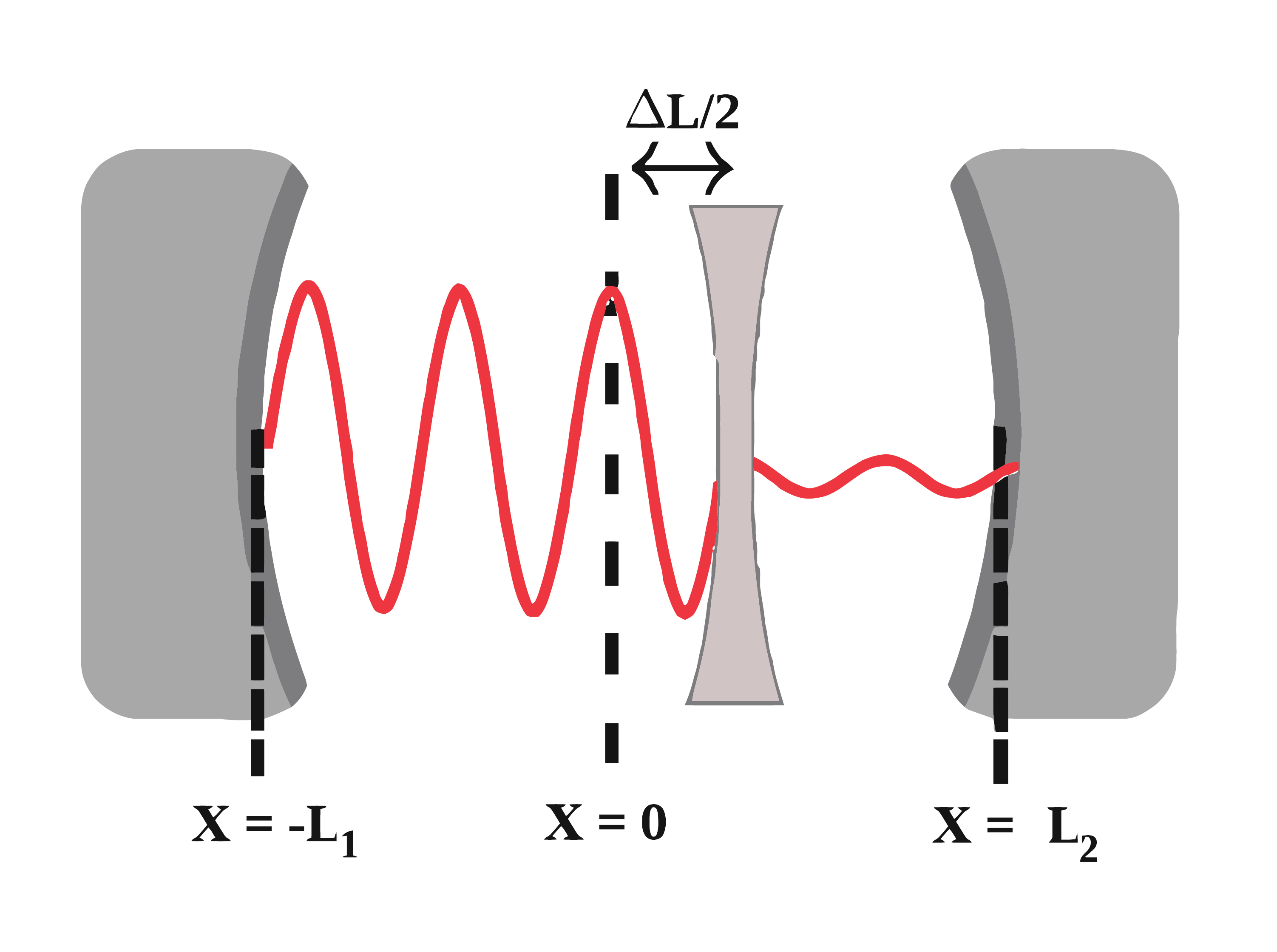}
\caption{(Color online) Double cavity setup consisting
of two perfectly reflecting mirrors separated by a partially transmissive common central mirror. $\Delta L \equiv L_{1}-L_{2}$ is the difference in length between the two cavities.}
\label{fig:cavitysetup}
\end{figure}

In this paper we consider the adiabatic transfer of light between two coupled cavities by changing their lengths. This can be achieved, for example, by translating the cavity mirrors using piezo-electric actuators (motors). Alternatively, if the cavity takes the form of a dielectric waveguide the mirror positions can remain fixed but the optical length of the cavities can be controlled by modulating the refractive index \cite{preble07}.  The specific model system we choose to illustrate the adiabatic transfer process in either of these experimental scenarios is a double cavity in one dimension as depicted in Figure \ref{fig:cavitysetup}. Changing the lengths of the two cavities changes the structure of the electromagnetic modes in such a way that a mode which is initially localized in one cavity can be smoothly transferred into the other cavity.  Our study of the spatial adiabatic passage of light is in the same spirit as Pelizzari's, although we shall not consider the atoms, but focus instead on the global normal modes of Maxwell's equations whose frequencies form a net of avoided crossings as a function of mirror displacement. It would be straightforward to introduce atoms into the scheme: for certain mirror displacements a quasi-resonant global light mode can be strongly localized in one of the cavities, and so a photon emitted by an atom in that cavity will find itself in that chosen mode. This global mode is then transferred to the other cavity, where the other atom can interact with the photon with high efficiency. On the other hand, if during the emission the mirror is not at a position which localizes the relevant global mode in the first cavity, then the photon will be placed into a superposition of global modes and the resulting time-dependence of the superposition leads to an oscillation of the photon back and forth between the two cavities, thereby lowering the transfer efficiency to the other atom and/or introducing the necessity for good timing.

Because our model is simple, it allows  a straightforward analysis of the conditions for adiabatic transfer, and a number of analytic results can be obtained. Furthermore, the Yale opto-mechanical group have realized a double cavity \cite{thompson08,jayich08} which is close to that considered by us here. Their experimental setup consists of an optical cavity divided in half by a thin silicon membrane. When the cavity is driven through an end mirror by an external laser, the membrane vibrates due to radiation pressure. Indeed, a recent theoretical paper  \cite{heinrich10}, discusses photon shuttling between the two halves of the cavity, in close correspondence to what we will consider here. Other theoretical studies of this system treat it as a photonic version of the Josephson junction \cite{chefles95,gerace09,larson10}. The important difference between our paper and previous optomechanical studies is that we propose to control the motion of mirror from outside rather than allowing it to move under the action of radiation pressure.

The problem of a cavity with moving mirrors is also of general interest from the point of view of studying CQED with time-dependent boundary conditions. This is an interesting generalization of CQED with many potential applications, including the field of opto-mechanics.  Other related applications include the possibility of modifying the decay rate of light from a cavity by modulating the position of the mirrors in time \cite{liningtonthesis,linington06,linington09}, and if the mirrors accelerate very rapidly, photon pair production \cite{eberlein,lambrecht05,johansson09}. Observations of this dynamical Casimir effect have recently been reported in microwave cavities \cite{wilson10}.  Also on the experimental side, advances in the ultrafast manipulation of the refractive index of silicon microcavities (which is equivalent to changing the cavity's length) have been used to demonstrate transitions between modes \cite{dong08}. When the refractive index is changed more slowly, the light adiabatically follows the modes, leading to a change in its color \cite{preble07,reed03,yanik04,xu07}.

The programme we follow in this paper is to treat cavity electrodynamics with time-dependent boundary conditions from first principles, by starting with Maxwell's wave equation. From this we derive an effective `hamiltonian'  which can be used for the purposes of first-order time propagation.  More precisely, by making an approximation analogous to the paraxial approximation familiar from optics, but in time rather than space, we find that the Maxwell wave equation reduces to an equation of motion for the classical Maxwell field that is first order in time and has a mathematical resemblance to the time-dependent Schr\"{o}dinger equation \cite{liningtonthesis}. (For a different approach to mapping the Maxwell wave equation onto the Schr\"{o}dinger equation see \cite{winn99,malkova03}). Under certain conditions we are therefore able to apply Landau-Zener theory \cite{landau32,zener32,stenholm96}, familiar from non-relativistic quantum mechanics, to the calculation of the coupling between classical Maxwell modes induced by changing the lengths of the cavities. We emphasize that our method is different from one in which the hamiltonian obtained from the \emph{static} problem is used for time propagation.  In particular, we show that there are higher order terms which provide significant corrections to the latter scheme \cite{liningtonthesis}.

The idea of using the Landau-Zener model to understand time-dependent processes in optics is not new. For example, the experiment reported in \cite{bouwmeester95} studied the time evolution of light in cavity where an electo-optic modulator was used to couple the two polarization states of a single longitudinal mode. A second electo-optic modulator  was used to vary the energy separation between two polarization states: varying this separation linearly in time whilst maintaining a constant coupling corresponds to the Landau-Zener model. Another system which is useful for studying the analogies between optical and quantum dynamics is provided by evanescently coupled waveguides, where the coupling between the waveguides changes with distance along the waveguide \cite{longhi07}. This system has been studied experimentally by a number of groups and various optical analogues to quantum phenomena have been observed, including Zener tunneling \cite{trompeter06}, adiabatic evolution \cite{lahini08}, and dynamical localization \cite{szameit10}. There have also been numerous studies of the conditions necessary for adiabaticity in CQED when atoms are included, see, for example, \cite{larson03}. In \cite{keeling08}, the dynamics of a two-level system coupled to a cavity field is studied in the situation where the separation between the two energy levels is varied linearly in time, and in \cite{mattinson01} the principle of STIRAP is applied to a two-mode Jaynes-Cummings model with degenerate mode frequencies, with the aim of adiabatically transferring photons between modes. 
We show in this paper that the application of the Landau-Zener model to the problem of a double cavity with changing cavity lengths illustrates in a simple fashion some of the similarities and differences between classical electrodynamics obeying Maxwell's equations and quantum mechanics obeying Schr\"{o}dinger's equation.

This paper is organized as follows. In Section \ref{sec:deltafunctionmirrormodel} we introduce a simple model for the spatial dependence of the dielectric permittivity function inside a double cavity. This model treats the central mirror as a Dirac $\delta$-function which facilitates analytic calculations later in the paper. In Section \ref{sec:MaxwellEqns} we find the global static solutions (normal modes) of Maxwell's wave equation subject to this dielectric function.  In Section \ref{sec:transferratio} we use these solutions to compute the transfer ratio, i.e.\ the degree to which the light can be localized on one side of the central mirror or the other
depending on mirror position. In Section \ref{sec:LZ} we introduce time-dependence by relating the problem of the transfer of light in a double cavity with a moving mirror to the well known Landau-Zener problem in quantum mechanics. In order to apply the results from the Landau-Zener problem we need to map the Maxwell wave equation, which is second order in time,  onto the  Schr\"{o}dinger wave equation, which is first order in time, and this is accomplished in Section \ref{sec:frommaxwelltoschrodinger}. The electric field is quantized in Section \ref{sec:quantization}, and in Section \ref{sec:regimes} we examine the regimes of validity of the mapping onto the  Schr\"{o}dinger wave equation. The $\delta$-function mirror model is compared to a more realistic model for a mirror in Section \ref{sec:FiniteWidthMirror}, and in Section  \ref{sec:experimentalfeasibility} we discuss the experimental feasibility. Conclusions are drawn in Section \ref{sec:conclusion}. 

This paper has six appendices: in Appendix \ref{sec:appendix1} we compare and contrast the double cavity problem for light with the double well problem in quantum mechanics. In Appendix \ref{sec:appendix2} we give some results on the transmission amplitudes for finite and $\delta$-function mirrors, and in Appendix \ref{sec:appendix2b} we explain how the results given in this paper can be applied to the case of two coupled waveguides with a fixed mirror position but two refractive indices, $n_{1}(t)$ and $n_{2}(t)$, that can be separately varied in time. In Appendix \ref{sec:appendix3} we define the diabatic basis and explain its connection to the adiabatic (global) basis. In Appendix \ref{sec:appendix4} we discuss the approximation whereby the time-dependence of diabatic basis functions is neglected. In Appendix \ref{sec:appendix5} we briefly consider modifications to Maxwell's wave equation due to  moving dielectric mirrors.

\section{$\delta$-function mirror model}
\label{sec:deltafunctionmirrormodel}

Consider a double cavity formed from two end mirrors plus a common mirror located between them, as shown Figure \ref{fig:cavitysetup}. We assume that the end mirrors are much more reflective than the central mirror and that the loss of light from the double cavity is negligible during the time taken for the light to be transferred from one side to the other. The stringent conditions this places on an experiment will be examined in Section \ref{sec:experimentalfeasibility}. As mentioned above, our proposed setup is motivated by remarkable developments in opto-mechanics \cite{thompson08,jayich08}. The additional feature we require here is that the common mirror be externally controlled, i.e.\ can be moved along the cavity axis. 

A simple theoretical model describing a double cavity has been given in a classic paper by 
Lang, Scully and Lamb \cite{lang73}. For the purposes of solving Maxwell's wave equation in the double cavity, they treated the end mirrors as perfect reflectors and the central mirror as a thin slab of dielectric material which is modelled by a Dirac $\delta$-function spatial profile. The double cavity model is thereby encoded in a dielectric permittivity function of the form
 \begin{equation}
\varepsilon(x)=\begin{cases}
\varepsilon_{0}(1+\alpha \delta(x)) & -L_{1}<x<L_{2}\\
\infty & \mbox{elsewhere}\end{cases}
\label{perm}
\end{equation}
where $x=-L_{1}$, and $x=L_{2}$ are the positions of the end mirrors. $\alpha$ is a parameter which determines the reflectivity of the common mirror, see the Appendices for more details. The total length of the double cavity is $L \equiv L_{1}+L_{2}$, and we also define the difference between the lengths of the two cavities to be $\Delta L \equiv L_{1}-L_{2}$,  which is also twice the displacement of the common mirror from the center of the whole cavity.

Maxwell's wave equation for the electric field $\mathcal{E}(x,t)$ in the double cavity is
\begin{equation}
\frac{\partial^{2}\mathcal{E}(x,t)}{\partial x^{2}}-\mu_{0}\varepsilon_{0}(1+\alpha\delta(x))\frac{\partial^{2}\mathcal{E}(x,t)}{\partial t^{2}}=0 \ .
\label{maxwell}
\end{equation}
We use this $\delta$-mirror model for the bulk of this paper because its simplicity facilitates analytic results. However, in Section \ref{sec:FiniteWidthMirror} we compare the results of the $\delta$-mirror model to the more realistic case of a mirror of finite width. 

Note that for mathematical convenience we have chosen a coordinate system in Eq.\ (\ref{perm}) where the common mirror is always located at $x=0$. However, we do not intend to attach any physical significance to this choice: physically, one can either take the view that the common mirror has a fixed position and it is the two end mirrors that are displaced, or vice versa. The latter view corresponds to an experimental situation, such as that described in \cite{thompson08} and \cite{jayich08}, where a common membrane is displaced, whilst the former view is closer in spirit to an experiment in which the position of the common mirror is held fixed, but the refractive index of the two cavities is modulated \cite{preble07}, see Appendix \ref{sec:appendix2b} for more details.  In Sections \ref{sec:LZ} and \ref{sec:frommaxwelltoschrodinger} we treat the case of moving mirrors: providing the mirrors are moving at constant velocities, it is physically equivalent to imagine either translating the end mirrors or the common mirror because these two situations simply correspond to two different inertial frames. However, there are some interesting considerations to be taken into account when treating light propagating in moving dielectrics, as briefly discussed in Appendix \ref{sec:appendix5}.

\section{Solutions of Maxwell's Wave Equation}
\label{sec:MaxwellEqns}

We write the solutions to the Maxwell wave equation as $\mathcal{E}_{m}(x,t)=U_{m}(x) \exp(-i\omega_{m}t)$,  where $\omega_{m}=k_{m}/\sqrt{\varepsilon_{0}\mu_{0}}$ is the angular frequency and $m=1,2,3 \ldots$ is an integer labelling the modes. The dimensionless mode functions $U_{m}(x)$ can be chosen to be orthonormal
in the Sturm-Liouville sense by ensuring that they obey
\begin{equation}
\frac{1}{\varepsilon_{0}}\int_{-L_{1}}^{L_{2}}\varepsilon(x)U_{l}(x)U_{m}(x)dx=\delta_{lm} \ .
\label{normalization}
\end{equation}
Inserting the above form for $\mathcal{E}(x,t)$ into Eq.\ (\ref{maxwell}) gives
\begin{equation}
\frac{\mathrm{d}^{2}U_{m}(x)}{\mathrm{d}x^{2}}+k_{m}^{2}(1+\alpha\delta(x))U_{m}(x)=0 \ .
\label{maxwell2}
\end{equation}
Solutions satisfying the boundary conditions $U_{m}(-L_{1})=U_{m}(L_{2})=0$
are given by
\begin{equation}
U_{m}(x)=\begin{cases}
A_{m}\sin \left[k_{m}(x+L_{1})\right]\quad & -L_{1} \leq x\leq0\\
B_{m}\sin \left[k_{m}(x-L_{2})\right]\quad & \,\:\:\:0 \leq x \leq L_{2} \ . \end{cases}
\label{Wavemode}
\end{equation}
One boundary condition is given by the continuity of the electric field across
the $\delta$-mirror, $U_{m}(0^{+})=U_{m}(0^{-})$. A second boundary condition is given by integrating Eq.\ (\ref{maxwell2}) over a vanishingly small interval containing the mirror, leading to $U_{m}^{\prime}(0^{+})-U_{m}^{\prime}(0^{-})=-\alpha k_{m}^{2}U_{m}(0)$.

\begin{figure}
\includegraphics[width=1.0\columnwidth]{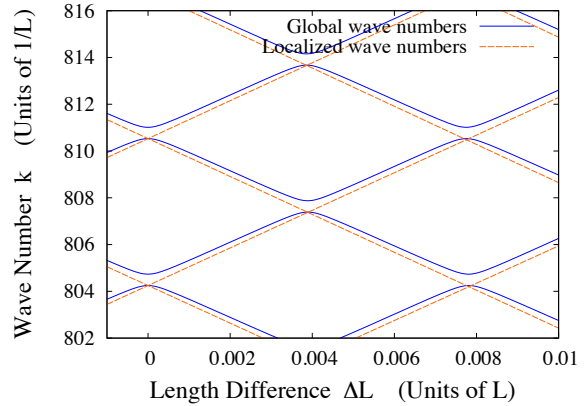}
\caption{(Color online) The wave numbers $k$ of the modes in a double cavity plotted as a function of $\Delta L \equiv L_{1}-L_{2}$ which is the difference in length between the two cavities.  The wave numbers are obtained by solving the transcendental equation (\ref{transcendental2}). When the reflectivity of the common mirror is infinite ($\alpha \rightarrow \infty$),  we obtain the localized modes (dashed red lines), which are trapped on one side or the other of the  mirror. When the reflectivity is finite we obtain the global modes (solid blue curves) which feature avoided crossings.  The total cavity length is set at $L=1 \times 10^{-4}$m, and  the global curves have $\alpha = 1 \times 10^{-6}$m, which translates to a transmission probability of the common mirror  of about 4\%. }
\label{fig:avoidedcrossingnet}
\end{figure}

Combining the two boundary conditions one is led to the following equation
for the wave numbers $k_{m}$ of the allowed modes \cite{lang73}
\begin{equation}
\tan(k_{m} L_{2})=\frac{\tan(k_{m} L_{1})}{\alpha k_{m} \tan(k_{m} L_{1})-1} \ .
\label{transcendental1}
\end{equation}
This transcendental equation can in general only be solved numerically. It is useful to re-write it as
\begin{equation}
\cos(k_{m} \Delta L)-\cos(k_{m} L)=2 \ \frac{\sin k_{m} L}{\alpha k_{m}} \, ,
\label{transcendental2}
\end{equation}
so that when $\alpha k$ is large the sinc function on the right hand side is small. The left hand side may then be expanded around its roots and this permits approximate analytic solutions (Eqns (\ref{even approx}) and (\ref{odd approx}) below).

\begin{figure}
\includegraphics[width=1.0\columnwidth]{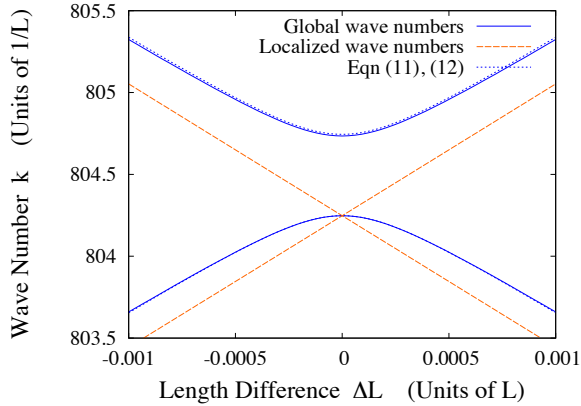}
\caption{(Color online) A zoom-in of one of the avoided
crossings shown in Figure \ref{fig:avoidedcrossingnet} corresponding to $n=128$. Three pairs of curves are shown:
the exact numerical solutions of Eq.\ (\ref{transcendental2}) are plotted as solid (blue) curves, the approximate analytical solutions given by Eqns (\ref{even approx2}) and (\ref{odd approx2}) are plotted as short dashed (black) curves, and the localized solutions are plotted as long dashed (red) curves. The difference between the exact numerical solutions and the approximate analytical ones is hard to discern.
Note that in contrast to the double well problem in quantum mechanics, the even mode lies at a higher  frequency $\omega=ck$ than the odd mode.}
\label{fig:avoidedcrossing}
\end{figure}

Consider first the situation when the common mirror is perfectly reflecting ($\alpha \rightarrow \infty$), so that the two sides of the double cavity are uncoupled. The modes in this case are specified by the wave numbers $k_{n}=2 n \pi/ (L\pm \Delta L)$, which come in pairs  that cross at $\Delta L=0$, and are plotted as the dashed red lines in Figures \ref{fig:avoidedcrossingnet} and \ref{fig:avoidedcrossing}. The mode whose wave number decreases with increasing $\Delta L$ corresponds to light trapped on the left hand side of the mirror, and vice versa. 
When the modes are coupled there are still two for each value of the integer $n$, whereas each value of the integer $m$ introduced previously labels a single mode: from henceforth we shall use the ``$n$'' convention. 

We refer to the solutions of the wave equation in the perfectly reflecting case as \emph{localized} modes.
To find the modes when the coupling between the two sides of the cavity is finite, one must solve Eq.\ (\ref{transcendental2}) for $k$, with $\alpha$ appropriately chosen to give the desired reflectivity of the common mirror. As shown in Appendix \ref{sec:appendix2}, the reflection probability from the $\delta$-mirror is given by
\begin{equation}
R  = \frac{k^2 \alpha^2 /4}{1+k^2 \alpha^2 /4} \ .
\label{eq:deltareflectivity}
\end{equation}
We shall refer to solutions of Eq.\ (\ref{transcendental2}) for finite $\alpha$ as \emph{global} solutions, because they extend throughout the double cavity, and they are plotted as the solid curves in Figures \ref{fig:avoidedcrossingnet} and \ref{fig:avoidedcrossing}. As can be seen in these figures, the effect of the coupling is to turn crossings into avoided crossings.

At $\Delta L =0$ the cavity is symmetric about the central mirror and so the solutions of Maxwell's wave equation must have well defined parity: for each $n$ we find a pair of modes, one of which is even and the other odd. As the common mirror is moved away from the center of the cavity the global modes lose their well defined parity, but we shall continue to label them as ``even'' and ``odd'' for the sake of simplicity.  By expanding Eq.\  (\ref{transcendental2}) to second order in $\Delta L$ about $\Delta{L}=0$, we obtain the following expressions for the wave numbers of the global modes valid close to avoided crossings centered at $\Delta L=0$
\begin{eqnarray}
k_{n,e} & \approx & \frac{2\pi n}{L}+\frac{2 \pi n}{L}\frac{1}{1+n^{2} \pi^2 \alpha/L}+\frac{2\pi^{3}n^{3}\alpha \Delta L^{2}}{L^4}\label{even approx} \\
k_{n,o} & \approx & \frac{2\pi n}{L}- \frac{2\pi^{3}n^{3}\alpha \Delta L^{2}}{L^4}\label{odd approx} \ .
\end{eqnarray}
These analytic approximations give very good agreement to the exact numerical solutions of Eq.\  (\ref{transcendental2}) provided one is quite high up in the spectrum (optical regime), i.e.\ when $n \gg 1$, but are only valid in the immediate vicinity of an avoided crossing. In particular, being quadratic they do not display the asymptotically linear behavior away from the avoided crossing that can be seen in Fig.\ \ref{fig:avoidedcrossing}. This can be remedied by fitting the wave numbers close to an avoided crossing to the functions
\begin{eqnarray}
k_{n,e} & = & \frac{2\pi n}{L} + \frac{\Delta}{\hbar c} + \frac{1}{\hbar c} \sqrt{\Delta^2+ \gamma \, \Delta L^2} \label{even approx2} \\
k_{n,o} & = & \frac{2\pi n}{L} + \frac{\Delta}{\hbar c} -  \frac{1}{\hbar c} \sqrt{\Delta^2+ \gamma \, \Delta L^2} \label{odd approx2}
\end{eqnarray} 
which are the forms suggested by the Landau-Zener hamiltonian (\ref{LZ-Hamiltonian}), as will be explained in  Section \ref{sec:LZ}. The gap between the wave numbers of the even and odd global solutions at an avoided crossing is given by $2 \Delta/\hbar c$, where for later convenience we have defined $\Delta$ as an energy.
The parameters $\gamma$ and $\Delta$ can be obtained by matching the Taylor expansions of Eqns (\ref{even approx2}) and (\ref{odd approx2}) to Eqns  (\ref{even approx}) and (\ref{odd approx}), respectively. We find
\begin{eqnarray}
\gamma & = & 2 \Delta \, \hbar c \,   \frac{2\pi^{3}n^{3}\alpha}{L^4} \label{eq:gamma} \\
\Delta & = & \frac{\hbar c}{L}\frac{n \pi}{1+n^2 \pi^2 \alpha/L  } \, . \label{Delta approx}
\end{eqnarray}
Note that this expression for $\Delta$ is valid up to second order in the quantity $\epsilon=kL-2 \pi n$, i.e.\ the deviation of the solutions from the perfectly localized case. $\epsilon$ is small if $\alpha k \gg 1$, which is the case for reasonably reflective mirrors and optical wave numbers. 
 A comparison between the values of $k$ found from an exact numerical solution of Eq.\  (\ref{transcendental2}), and the approximate analytic solutions given by Eqns (\ref{even approx2}) and (\ref{odd approx2}) is included in Fig.\ \ref{fig:avoidedcrossing}.

\begin{figure}
\includegraphics[width=1.0\columnwidth]{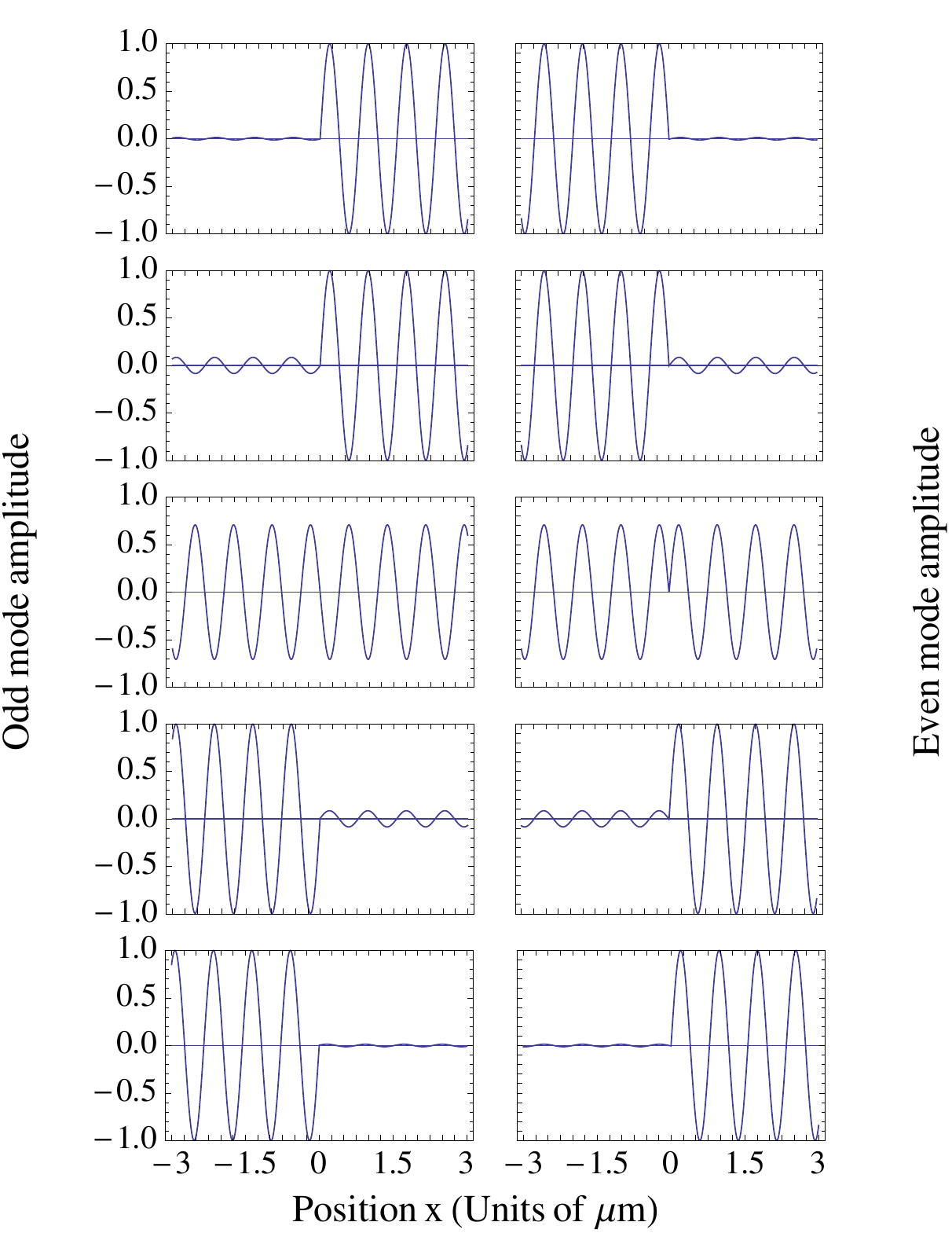}
\caption{This series of plots shows how each global wave mode swaps from one side of the double cavity to the other as the common mirror is moved through an avoided crossing. The odd mode $U_{n,\mathrm{odd}}(x)$ plotted as a function of position $x$ is shown in the left column, and the even mode $U_{n,\mathrm{even}}(x)$ is shown in the right column. Each row corresponds to a different value of  $\Delta L / L$: the top row has  $\Delta L / L= - 4 \times 10^{-4}$, the second row has  $\Delta L / L= - 0.6 \times 10^{-4}$, the middle row  has $\Delta L / L= 0$,  the fourth row has  $\Delta L / L= 0.6 \times 10^{-4}$, and the bottom row has  $\Delta L / L= 4 \times 10^{-4}$.  Strictly speaking, the modes only have well defined parity at $\Delta L=0$, but it is convenient to retain the labels ``even'' and ``odd'' for $\Delta L \neq 0$. The parameters used to make this figure were $L=1 \times 10^{-4}$m, $\alpha/L=0.3$, and $n=128$ so that
$k \approx 8 \times 10^6$m$^{-1}$.}
\label{fig:ModePlotNew}
\end{figure}

Away from the avoided crossings the global modes become localized, each member of a pair
localizing on a different side of the common mirror. The larger $\alpha$ is, the stronger this localization becomes (and the more the avoided crossings close up). The general procedure for the adiabatic transfer of a light mode from one side of the double cavity to the other is now apparent: if the common mirror position is initially set to one side of an avoided crossing (the optimal position is roughly halfway to the next avoided crossing--see Section \ref{sec:transferratio}) then the localized and global modes approximately coincide and the mode is localized on one side on the common mirror.  If the mirror is now slowly translated to the equivalent point on the other side of the avoided crossing, the system will follow the global curve on which it began. On the opposite side of the avoided crossing this global mode approximately coincides with the  localized mode on the other side of the common mirror. This procedure is illustrated in Figure \ref{fig:ModePlotNew}.

It is notable that the wave number of the degenerate localized solutions at $\Delta L=0$ coincides exactly with the lower branch of the corresponding global solutions at the avoided crossing, rather than lying halfway between the two global branches.  This is because the lower global branch corresponds to an odd solution in the double cavity. Odd solutions have a node at the $\delta$-mirror and hence do not see the mirror at all. Connoisseurs of the related problem of the double well potential in quantum mechanics \cite{landau&lifshitz} will immediately observe that this is the reverse of the situation found there, where the lower branch is due to the even global solution. This reversal  of the ordering of the solutions with respect to the double well problem in quantum mechanics is due to the fact that for light propagating in 1D there can be no evanescent waves (assuming the refractive index is $> 1$ and there is no absorption) and so there can be no tunnelling barriers, only wells. The double cavity system for classical light waves is actually equivalent to a quantum reflection problem in quantum mechanics. This point is discussed further in Appendix \ref{sec:appendix1}. 

\section{Transfer Ratio}
\label{sec:transferratio}

\begin{figure}
\includegraphics[width=1.0\columnwidth]{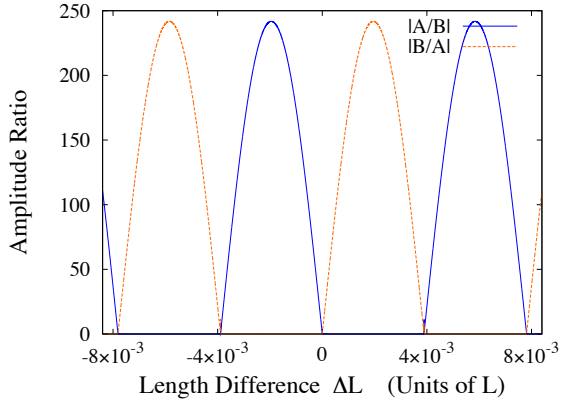}
\caption{(Color online) The relative amplitudes $A_{n}/B_{n}$
and $B_{n}/A_{n}$ of the modes as a function of the displacement $\Delta L$ of the common mirror. The parameters used to make this plot were $\alpha/L=0.3$,  and $n=128$.}
\label{fig:AoverB}
\end{figure}

Once the wave numbers have been determined, the field amplitudes $A_{n}^{o,e}$ and $B_{n}^{o,e}$ on the two sides of the common mirror can be calculated.  From the continuity condition for the field across the mirror we find that   
\begin{equation}
\frac{A_{n}}{B_{n}}=-\frac{\sin(k_{n}L_{2})}{\sin(k_{n}L_{1})} = -\frac{\sin [k_{n} (L-\Delta L)/2]}{\sin[k_{n}(L+\Delta L)/2]} \label{A/B} 
\end{equation}
where we have suppressed the even/odd labels.
Using this ratio one can construct the global field distribution inside the cavity, as shown in Figure \ref{fig:ModePlotNew} for a pair of modes. The mirror only needs to be moved by approximately one quarter of a wavelength to achieve the transfer of light from one cavity to the other.

The ratio given by Eq.\ (\ref{A/B}) is a key quantity because it measures the degree to which the field is localized on one side or the other.
In Figure \ref{fig:AoverB} we plot $A/B$ as a function of $\Delta L$, for the case
$\alpha/L=0.3$.  The maxima of the ratio $A/B$ are located at values of the mirror displacement which are roughly halfway between pairs of avoided crossings. We shall denote these values of the mirror displacement by $\Delta L_{n \star}$.  Accordingly, we shall denote the values of the wave numbers at these points by $k_{n \star}$. These wave numbers obey the equation
\begin{equation}
\cos (k_{n \star} L)+ \frac{\sin(k_{n \star} L)}{k_{n \star} L}\left( \frac{2+\alpha/L}{\alpha/L} \right)=0 \, ,
\label{eq:knstartrancendental}
\end{equation}
which can be obtained by finding the stationary points of the ratio (\ref{A/B}) with respect to $\Delta L$, and eliminating the quantity $\mathrm{d} k_{n \star}/ \mathrm{d} \Delta L$, which appears as a result of the differentiation, by finding an expression for $\mathrm{d} k_{n \star}/ \mathrm{d} \Delta L$ from Eq.\ (\ref{transcendental2}).

Equation (\ref{eq:knstartrancendental}) can be used to find an analytic expression for $k_{n \star}$.
In the regime $\alpha k \gg 1$, the wave numbers $k_{n \star}$ are located very close to the points halfway between the avoided crossings, $k_{n \star} \approx (4n \pm 1) \pi/(2L)$, where the $\pm$ sign refers to the even/odd mode. Expanding Eq.\ (\ref{eq:knstartrancendental}) about these points we find that to first order 
\begin{equation}
k_{n \star}L= (4n\pm1) \pi/2 +\frac{2+\alpha/L}{2 \pi n \alpha/L }
\, .
\label{x_0}
\end{equation}
The mirror displacement $\Delta L_{n \star}$ corresponding to wave number $k_{n \star}$ is found by solving Eq.\ (\ref{transcendental2}) for $\Delta L$ and evaluating it at $k_{n \star}$, i.e.\
\begin{eqnarray}
\Delta L_{n \star} & = & \frac{1}{k_{n \star}}\arccos \left[\frac{2\sin k_{n \star}L}{\alpha k_{n \star}}+\cos k_{n \star} L \right] 
 \\
& \approx & \frac{1}{k_{n\star}}\left(\frac{\pi}{2}
-\frac{2}{\alpha k_{n\star}}+k_{n\star}L-(4n\pm1)\frac{\pi}{2}\right) \label{L_0}
\end{eqnarray}
where to obtain the  second line we expanded the $\arccos$ term under the assumption that $\alpha k_{n \star} \gg 1$, and also made the approximation $\sin (k_{n \star} L) \approx 1$.

\begin{figure}
\includegraphics[width=1.0\columnwidth]{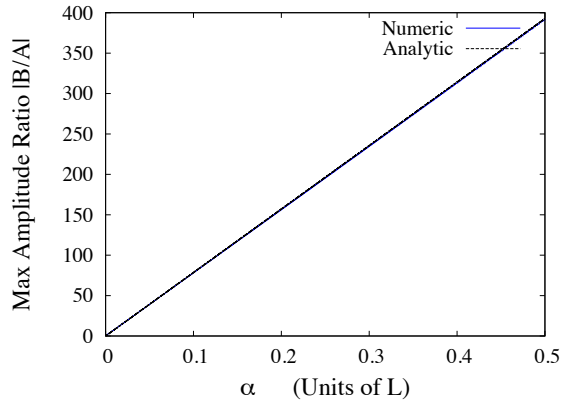}
\caption{(Color online) The absolute value of the maximum value the relative amplitude $B_{n}/A_{n}$ can take as a function of the common mirror parameter $\alpha$. The exact numerical result is plotted as the solid curve, and the approximate analytical result, as given by Eq.\ (\ref{eq:maxA/Bapprox}), is plotted as the dashed curve. Note that the difference between these two curves is hard to discern.  We set n=128 to make this plot.}
\label{fig:MaxAoverB}
\end{figure}

Having found expressions for $k_{n \star}$ and $\Delta L_{n \star}$, we are now in a position to obtain analytic estimates for the maximum value of the transfer ratio given in Eq.\ (\ref{A/B}). This ratio can first be simplified by noting that at a maximum we expect the numerator  to be near unity and the denominator to be near zero. Since it is  the inverse ratio $B/A$ which is large for $\Delta L$ small and positive, let us consider that. Expanding the numerator about the point where the phase is equal to an odd integer multiple of $\pi/2$ gives
\begin{equation}
\sin\left(\frac{k_{n}}{2}(L+\Delta L_{n})\right) \approx 1-\frac{1}{2}\left(\frac{k_{n}}{2}(L+\Delta L_{n})-\frac{\pi}{2}-\pi n\right)^{2}
\end{equation}
and expanding the denominator about the nearest zero gives
\begin{equation}
\sin\left(\frac{k_{n}}{2}(L-\Delta L_{n})\right) \approx \frac{1}{2}k_{n}(L-\Delta L_{n})-\pi n
\end{equation}
so that close to a maximum the amplitude ratio can be written 
\begin{equation}
\left. \frac{B_{n}}{A_{n}}\right\vert_{\mathrm{max}} \approx \frac{-1+\frac{1}{8}\left(k_{n \star}(L+\Delta L_{n \star})-(\pi+2n\pi)\right)^{2}}{\frac{1}{2}k_{n \star}(L-\Delta L_{n \star})-n\pi} \, .
\end{equation}
Upon inserting Eqns (\ref{x_0}) and (\ref{L_0}) into this equation we still obtain a rather complicated result. However, if this expression is expanded for large values of $n$ and $k \alpha$ it simplifies to
\begin{equation}
\left. \frac{B_{n}}{A_{n}}  \right\vert_{\mathrm{max}}  \approx  - \frac{2n\pi}{L}\alpha \, .
\label{eq:maxA/Bapprox}
\end{equation}
This result is compared against the exact numerical result in Fig.\ \ref{fig:MaxAoverB}. The agreement is excellent, except when $\alpha/L$ becomes extremely small, in which case $\alpha k_{n \star} \approx 2 n \pi \alpha /L$ is no longer large, and the approximations made in the above derivation break down. 

As can be seen in Fig.\ \ref{fig:MaxAoverB}, as $\alpha$ increases (i.e.\ as the reflectivity of the common mirror increases), so too does the maximum possible shift in amplitude. However, there is a trade-off: higher reflectivity gives smaller gaps at the avoided crossings, which requires slower mirror speeds if the transfer is to be adiabatic. This is discussed in Section \ref{sec:LZ}. 
Furthermore, whilst it is important to know the maximum value $A_{n}/B_{n}$ can take in order to evaluate the optimal performance of the double cavity as a light transfer device,  it is also useful to 
 know the ratio near $\Delta L=0$.  The reason for this is that, depending on the parameters, a sizeable amplitude ratio can already be reached at very small displacements, see Figure \ref{fig:AoverB}.  In practice, therefore, it may not be worthwhile going to the maximum, especially if that involves a longer transfer time. This issue is discussed further in Section \ref{sec:experimentalfeasibility}.  Using Eq.\ (\ref{even approx}) or (\ref{odd approx}) in the ratio (\ref{A/B}) yields the desired result.  This permits a very quick calculation of amplitude ratios near the equilibrium position, which along with the results of Section \ref{sec:LZ} may be used to minimize losses.

\section{Time-dependence: the Landau-Zener problem}
\label{sec:LZ}

So far we have found the time-independent solutions to the double cavity problem: the global modes are the static solutions to Maxwell's wave equation for a fixed mirror displacement.
Even if the system begins in a single (global) mode, a mirror moving at a finite speed will induce transitions and the light will therefore be excited into a superposition of modes. This is a complicated problem to treat analytically, but if we can map the passage of our system through an avoided crossing onto the related Landau-Zener (LZ) problem in quantum mechanics, then we can take advantage of existing solutions. 

The LZ problem considers a quantum system with two states, the energy separation of which is varied linearly in time so that at $t=0$ they are degenerate. The two states have a time-independent coupling $\Delta$. Stated in this way, the two states being coupled are known as the \emph{diabatic} states, and the time-dependent hamiltonian in the diabatic basis is
\begin{equation}
H_{LZ}=\left[\begin{array}{cc}
E(t) & \Delta\\
\Delta & -E(t) \end{array}\right] \, ,
\label{LZ-Hamiltonian}
\end{equation}
where the diabatic energies $\pm E(t)$ vary linearly in time: 
\begin{equation}
E = \vartheta t/2 \, . 
\end{equation}
Diagonalizing this hamiltonian at each instant of time gives the energies $E_{\pm}$ of the \emph{adiabatic} states 
\begin{equation}
E_{\pm}(t)=\pm \sqrt{\vartheta^2 t^2/4 + \Delta^2}
\end{equation}
which form an avoided crossing. At the point of closest approach ($t=0$) the magnitude of the gap between the two adiabatic energies is $2 \Delta$, and  at $t=\pm \infty$ the diabatic and adiabatic energies coincide.

The adiabatic states interchange their character between $t=-\infty$ and $t=+\infty$, which is the crucial property we need in order to transfer modes between the two sides of the double cavity. The exact solution of the LZ problem shows that if at $t = - \infty$ the two state system is prepared in one of the two states, then the probability that it has made a transition to the other adiabatic state by $t=+\infty$ is  \cite{landau32,zener32,stenholm96} 
\begin{equation}
P_{\mathrm{LZ}}=\exp(-  2 \pi \Delta^2/ \hbar \vartheta ) \ .
\label{eq:LZprobability}
\end{equation}
Of course, the LZ problem assumes that there are only two modes present, but due to the exponential suppression in the adiabatic regime of transitions as a function their energy difference, we need only consider a single even and odd pair in the double cavity if we start from a singly occupied mode and move slowly through a single avoided crossing.    

It is tempting to apply the result (\ref{eq:LZprobability}) directly to the time-dependent cavity problem by identifying the global modes we found in previous sections as the adiabatic states
\begin{eqnarray}
E_{+} & \equiv & \hbar \, c \, k_{e} \label{eq:E+=keven} \\
E_{-} & \equiv & \hbar \, c \, k_{o}  \label{eq:E-=kodd}
\end{eqnarray}
and extracting from them the parameters $\Delta$ and $\vartheta$ that are needed to calculate the transition probability $P_{\mathrm{LZ}}$. Note that the appearance of Planck's constant in these relations is purely for dimensional purposes (we have not yet quantized the electromagnetic field, see Appendix \ref{sec:appendix1} for more discussion).
In fact, the parameter $\Delta$ has already been found above and is given in Eq.\ (\ref{Delta approx}).
Furthermore, if the common mirror is moved at a constant velocity $v$, so that $\Delta L = 2 vt$, then the key requirement of the LZ model that the energy difference between the diabatic states is varied linearly with time will be fulfilled.  From Eqns (\ref{even approx2})--(\ref{Delta approx}) we find that $\vartheta= 4 v \sqrt{\gamma}$. For large $n$ this gives
\begin{equation}
\vartheta \approx 8 \pi \hbar c n  v/L^2 
\label{eq:thetaapprox}
\end{equation}
which coincides with the result obtained by assuming that the wave numbers are those given by localized solutions.

However, this whole approach  of directly applying the LZ results to the time-independent solutions of the Maxwell equations is wrong in principle because the Maxwell and Schr\"{o}dinger equations evolve differently in time. Therefore, although the time-independent Maxwell and Schr\"{o}dinger equations are identical (compare Eqns (\ref{eq:maxwelltimeindependent}) and (\ref{eq:schrodingertimeindependent})), one should not blindly apply the LZ solution to the Maxwell problem. Rather, in Section \ref{sec:frommaxwelltoschrodinger} we justify the application of the LZ result to the moving mirror problem by deriving an approximate version of the Maxwell equation which is first order in time and is
valid in the adiabatic regime. In order to set this up, we need to express our system in the LZ diabatic basis. Although they are similar in character, the diabatic states do not correspond exactly to the localized modes unless $\alpha \rightarrow \infty$ (the LZ diabatic energies must cross halfway between the two adiabatic curves whereas the localized modes do not). There is a precise way to obtain the LZ diabatic modes and this ``undiagonalization'' is presented in Appendix \ref{sec:appendix3}. 

The reason it is advantageous to work in the diabatic basis is worth mentioning. Consider the time-dependent solution expressed in the diabatic and adiabatic bases. Denoting the diabatic modes as $\phi_{L}(x)$ and $\phi_{R}(x)$, we have 
\begin{eqnarray}
\mathcal{E}(x,t) & = & A_{L}(t)\phi_{L}(x,t)+A_{R}(t)\phi_{R}(x,t) \label{eq:diabatic expansion} \\
\mathcal{E}(x,t) & = & B_{e}(t) U_{e}(x,t)+B_{o}(t) U_{o}(x,t) \ .
\end{eqnarray}
The time-dependence extends not only to the superposition coefficients but also to the basis modes themselves and this complicates the treatment. As can be seen \emph{a posteriori} from the LZ transition probability (\ref{eq:LZprobability}), the changes in the adiabatic superposition coefficients $B_{e}(t)$ and $B_{o}(t)$ are exponentially small during a transfer in the adiabatic regime. Almost all the change is incorporated in the adiabatic modes.
By contrast, the changes in the diabatic coefficients during a transfer in the adiabatic regime are of order unity because in this basis the system changes its state. As pointed out by Zener himself \cite{zener32}, the changes in the diabatic basis in the transition region (close to the avoided crossing) are typically negligible in comparison to the changes in the coefficients, and so one can take $\phi_{L}(x)$ and $\phi_{R}(x)$ to be independent of time. This approximation is analyzed in Appendix \ref{sec:appendix4}. Thus, by working in the diabatic basis  we can focus purely on the time-dependence of the coefficients $A_{L}(t)$ and $A_{R}(t)$, which is a significant simplification.

\section{From Maxwell to Schr\"{o}dinger}
\label{sec:frommaxwelltoschrodinger}

Our aim in this section is to derive from Maxwell's wave equation a Schr\"{o}dinger-like (first order in time) equation for the electric field amplitudes. This will allow us to put the use of the LZ result in the time-dependent double cavity problem on a firm footing.

As discussed above, it is more convenient to solve the LZ  problem in the diabatic basis. On the other hand, the exact solutions we have already found by solving Eq.\ (\ref{transcendental2}) are for the global modes, i.e.\ for the adiabatic basis. We shall therefore switch backwards and forwards between the two bases as needed. We begin in the diabatic basis and substitute expression (\ref{eq:diabatic expansion}) for the electric field into the Maxwell wave equation $\partial_{x}^{2} \mathcal{E} = \mu_{0} \varepsilon(x,t) \partial_{t}^{2} \mathcal{E}$, neglecting the time-dependence of the mode functions $\phi(x,t)$. The validity of this approximation is examined in Appendix \ref{sec:appendix4}.   This gives
\begin{eqnarray}
A_{L} \partial_{x}^{2} \phi_{L}+A_{R} \partial_{x}^2 \phi_{R} = & & \mu_{0} \varepsilon(x,t) \big(\ddot{A}_{L} \phi_{L} +\ddot{A}_{R}\phi_{R} \big) \label{eq:maxwell2b}
\end{eqnarray}
where the dots indicate time derivatives.   In fact, the time dependence of the dielectric function $\varepsilon(x,t)$ implies that there are also corrections to the above quoted form of Maxwell's wave equation \cite{leonhardt99}; these corrections are discussed in Appendix \ref{sec:appendix5}.

We tackle the left and right hand sides of Eq.\ (\ref{eq:maxwell2b}) separately.
To simplify the left hand side we substitute the expansions (\ref{phi-L}) and (\ref{phi-R}) of the localized modes in terms of the global modes and then make use of the fact that the global solutions, being eigenmodes of the system, satisfy
\begin{eqnarray}
\partial_{x}^2 U_{e}= - \mu_{0} \varepsilon(x,t) \omega_{e}^2 U_{e} \\
\partial_{x}^2 U_{o}= - \mu_{0} \varepsilon(x,t) \omega_{o}^2 U_{o}
\end{eqnarray}
to give
\begin{eqnarray}
\partial_{x}^2 \phi_{L} & = & - \mu_{0} \varepsilon(x,t) \left(-\sin (\theta) \omega_{e}^2 U_{e}+ \cos (\theta) \omega_{o}^2 U_{o}\right) \\
\partial_{x}^2 \phi_{R} & = & - \mu_{0} \varepsilon(x,t) \left(\cos (\theta) \omega_{e}^2 U_{e}+ \sin (\theta) \omega_{o}^2 U_{o}\right) \ .
\end{eqnarray}
Now that we have made use of the exact results for the global modes we can go back to the diabatic basis by using (\ref{eq:Ueven}) and (\ref{eq:Uodd}) to obtain
\begin{eqnarray}
\partial_{x}^2 \phi_{L} & = & - \mu_{0} \varepsilon(x,t) \Big\{ \phi_{L} \big[\sin^{2}(\theta)  \omega_{e}^2 + \cos^{2} (\theta) \omega_{o}^2 \big] \nonumber \\
& + & \phi_{R} \big[\sin(\theta) \cos (\theta) \omega_{o}^2 -\sin(\theta) \cos (\theta) \omega_{e}^2 \big] \Big\}  \label{eq:lhs1a}\\
\partial_{x}^2 \phi_{R} & = & - \mu_{0} \varepsilon(x,t) \Big\{ \phi_{R} \big[\cos^{2}(\theta)  \omega_{e}^2 + \sin^{2} (\theta) \omega_{o}^2 \big] \nonumber \\
& + & \phi_{L} \big[\sin(\theta) \cos (\theta) \omega_{o}^2 - \sin(\theta) \cos (\theta) \omega_{e}^2 \big] \Big\} \ . \label{eq:lhs1b}
\end{eqnarray}

In order to explicitly express the adiabatic transfer problem in LZ form we let
\begin{eqnarray}
\omega_{e} & = & \sqrt{E^{2}+\Delta^{2}}/\hbar+\omega_{\mathrm{av}}
\label{omega-even} \\
\omega_{o} & = &-\sqrt{E^{2}+\Delta^{2}}/\hbar+\omega_{\mathrm{av}} \ ,
\label{omega-odd}
\end{eqnarray}
cf. Eqns (\ref{even approx2}) and (\ref{odd approx2}).
In these equations $\pm E/\hbar$ are the frequencies of the diabatic modes, as discussed in Appendix \ref{sec:appendix3}. Indeed, Eqns (\ref{omega-even}) and (\ref{omega-odd}) can be viewed as a way of defining the diabatic modes, the adiabatic (global) modes with frequencies $\omega_{e}$ and $\omega_{o}$ being already known from solving the transcendental equation (\ref{transcendental2}).  We have also introduced $\omega_{\mathrm{av}}$ which is the average frequency of the even and odd modes 
\begin{equation}
 \omega_{\mathrm{av}} = \frac{\omega_{e}+\omega_{o}}{2}
\label{omega-average}
\end{equation}
and takes account of the fact that the avoided crossings are not located at frequency $E/\hbar=0$, but are in fact up in the optical spectrum. From Eqns (\ref{even approx2}) and (\ref{odd approx2}) we see that 
\begin{eqnarray} 
\omega_{\mathrm{av}} = \frac{2\pi n c}{L} + \frac{\Delta}{\hbar } \, .
\end{eqnarray}  
The definitions (\ref{omega-even}) and (\ref{omega-odd}) can be combined with expressions (\ref{cos}) and (\ref{sin}) for  $\cos (\theta)$ and $\sin (\theta)$ in terms of the diabatic variables to give
\begin{equation}
\sin (\theta) \cos (\theta) \left( \omega_{o}^2-\omega_{e}^2 \right)= 2 \Delta \omega_{\mathrm{av}}/\hbar  
\end{equation}
which is one of the combinations appearing in (\ref{eq:lhs1a}) and (\ref{eq:lhs1b}). In a similar fashion we find that the other combinations are given by
\begin{eqnarray}
 \sin^{2}(\theta) \omega_{e}^2+\cos^{2} (\theta) \omega_{o}^2 &=& (E/\hbar-\omega_{\mathrm{av}})^2+\Delta^2/\hbar^{2} \\
 \cos^{2}(\theta) \omega_{e}^2+\sin^{2} (\theta) \omega_{o}^2 &=& (E/\hbar+\omega_{\mathrm{av}})^2+\Delta^2/\hbar^{2} \ .
\end{eqnarray} 
Putting all this together, and combining with the right hand side of Eq.\ (\ref{eq:maxwell2b}), we arrive at the following expression for the two-mode Maxwell wave equations
\begin{eqnarray}
&&  2 (\Delta/\hbar) \omega_{\mathrm{av}} A_{L} \phi_{R} + \{(E/\hbar-\omega_{\mathrm{av}})^2+\Delta^{2}/\hbar^2\} A_{L} \phi_{L} \nonumber \\
&& + 2 (\Delta/\hbar) \omega_{\mathrm{av}} A_{R} \phi_{L}+ \{(E/\hbar+\omega_{\mathrm{av}})^2+\Delta^{2}/\hbar^2\}A_{R} \phi_{R}   \nonumber \\
& & = - \left(\ddot{A}_{L} \phi_{L}+\ddot{A}_{R} \phi_{R} \right) \ .
\label{eq:maxwell2c}
\end{eqnarray}
Multiplying through by $\varepsilon(x) \phi_{L}(x)$ or $\varepsilon(x) \phi_{R}(x)$ and integrating over space in order to make use of the orthogonality conditions (\ref{eq:normalizationphi1}) and (\ref{eq:normalizationphi2}), we can pick out the coupled equations of motion for the amplitudes
\begin{equation}
-\left(\begin{array}{c}\ddot{A}_{L}  \\ \ddot{A}_{R}\end{array}\right)=\left(\begin{array}{cc} (\frac{E}{\hbar}-\omega_{\mathrm{av}})^2+\frac{\Delta^{2}}{\hbar^2} & 2 \Delta \omega_{\mathrm{av}}/\hbar \\ 2 \Delta \omega_{\mathrm{av}}/\hbar & (\frac{E}{\hbar}+\omega_{\mathrm{av}})^2+\frac{\Delta^{2}}{\hbar^2} \end{array}\right)\left(\begin{array}{c} A_{L}  \\ A_{R}\end{array}\right)
\label{eq:maxwell3}
\end{equation}

The two-mode equations of motion (\ref{eq:maxwell3}) are second order in time. They are dominated by the fast evolution generated by the diagonal terms containing $\omega_{\mathrm{av}}$ which is an optical frequency of order $10^{15}\mathrm{s}^{-1}$. This frequency is far greater than the other frequencies in the problem and comes from the stationary solutions. We can remove the diagonal terms by defining 
\begin{equation}
A_{L/R}(t) \equiv \widetilde{A}_{L/R}(t)  \exp \left[-i  \int_{0}^{t} \beta_{L/R}(t') \mathrm{d} t' \right] \, ,
\label{eq:Atilde}
\end{equation}
where
\begin{equation}
\beta_{L/R}(t)=\sqrt{(E(t)/\hbar \mp \omega_{\mathrm{av}})^2+\Delta^{2}/\hbar^{2}} \ .
\end{equation}
Time derivatives of $A_{L/R}(t)$ then give rise to the terms 
\begin{eqnarray}
\ddot{A}_{L/R} & = &\big(\ddot{\tilde{A}}_{L/R}-2 i \beta_{L/R} \dot{\tilde{A}}_{L/R}- i \dot{\beta}_{L/R}\tilde{A}_{L/R}-\beta_{L/R}^{2}\tilde{A}_{L/R} \big) \nonumber \\
&& \quad \quad \times  \exp \left[-i  \int_{0}^{t} \beta_{L/R}(t') \mathrm{d} t' \right] \ .
\label{eq:ddotA}
\end{eqnarray}
By virtue of the fact that $\beta_{L}$ and $\beta_{R}$ are so much greater than the other frequencies,  we make a `paraxial approximation in time' (or equivalently, a slowly varying amplitude approximation) and drop the first and third terms on the right hand side of Eq.\ (\ref{eq:ddotA}), and thereby obtain an approximation to Maxwell's wave equation that is first order in time 
\begin{eqnarray}
i\frac{\mathrm{d} \tilde{A}_{L/R}}{\mathrm{d}t}=\frac{ \omega_{\mathrm{av}}\Delta }{\hbar \beta_{L/R}}\tilde{A}_{R/L} \exp \left[ \pm i\int_{0}^{t}(\beta_{L}-\beta_{R}) \mathrm{d}t'\right] \, . 
\end{eqnarray}
These equations can be further simplified by noting that
\begin{equation}
\beta_{L/R} = \omega_{\mathrm{av}}\left(1\mp \frac{E}{\hbar \omega_{\mathrm{av}}}+\frac{1}{2}\frac{\Delta^2}{\hbar^2 \omega_{\mathrm{av}}^2} +\ldots \right) 
\label{eq:betaexpansion}
\end{equation}
and
\begin{equation}
\beta_{R}-\beta_{L}=2E(1-\frac{1}{2}\frac{\Delta^{2}}{\hbar^2 \omega_{\mathrm{av}}^2}+\ldots)
\label{eq:betadifferenceexpansion}
\end{equation}
and truncating each of these expansions at their first term. In this way we arrive at
\begin{equation}
i\frac{\mathrm{d}}{\mathrm{d} \tau}\left(\begin{array}{c} \tilde{A}_{L}  \\  \tilde{A}_{R} \end{array}\right)=\left(\begin{array}{cc}0 & \mathrm{e}^{-i \tilde{\vartheta} \tau^2/2} \\ \mathrm{e}^{i \tilde{\vartheta} \tau^2 /2} & 0\end{array}\right)\left(\begin{array}{c}\tilde{A}_{L}  \\ \tilde{A}_{R}\end{array}\right)
\label{eq:schrod}
\end{equation}
where $\tau \equiv \Delta t / \hbar$ and $\tilde{\vartheta} \equiv \hbar \vartheta / \Delta^2 $ are both dimensionless quantities. These coupled equations are mathematically equivalent to the time-dependent Schr\"{o}dinger equation $i \mathrm{d} \psi / \mathrm{d}\tau = H(\tau) \psi$ with the LZ hamiltonian (\ref{LZ-Hamiltonian}), as can be seen by letting $\tilde{A}_{L} \rightarrow \tilde{A}_{L} \exp [-i \tilde{\vartheta} \tau^2/4]$ and $\tilde{A}_{R} \rightarrow \tilde{A}_{R} \exp [i \tilde{\vartheta} \tau^2/4]$. They therefore justify, within their range of validity, the application of the LZ result (\ref{eq:LZprobability}) to the problem of a slowly moving mirror in a double cavity. However, it is important to bear in mind that these equations come entirely from classical electrodynamics. 

\begin{figure}
\includegraphics[width=1.0\columnwidth]{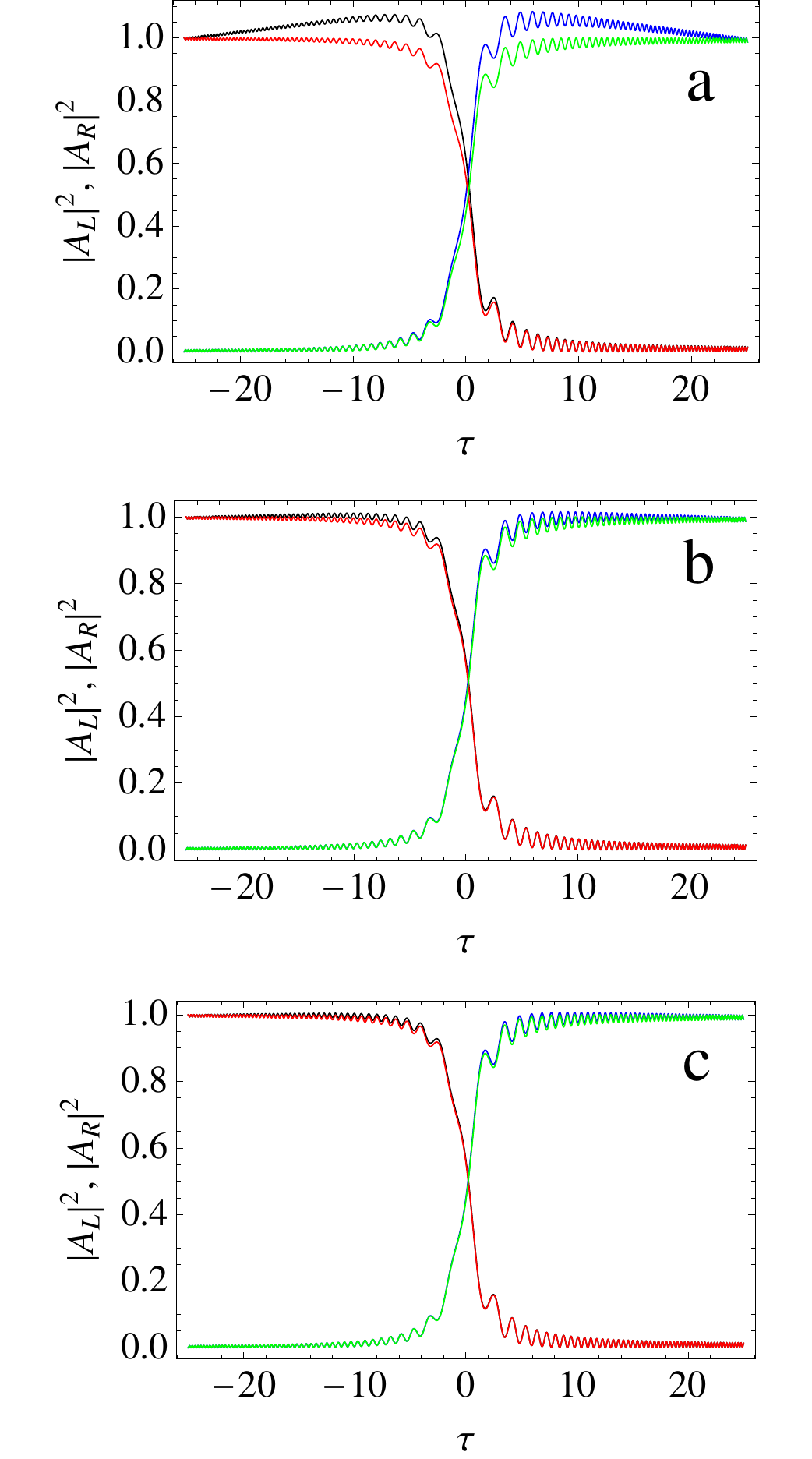}
\caption{(Color online) Comparison of second-order-in-time dynamics, as given by the full Maxwell wave equation (\ref{eq:maxwell3}), versus first-order-in-time dynamics, as given by the Schr\"{o}dinger-like equation (\ref{eq:schrod}), as the mirror is swept through an avoided crossing. The initial conditions are the same in every panel:  $A_{L}=1$ and $A_{R}=0$ at $\tau=-25$, where $\tau \equiv \Delta t / \hbar$ is dimensionless time. The first-order-in-time dynamics are also the same in every panel and are set by $\tilde{\vartheta} \equiv \hbar \vartheta/\Delta^2=1$, which is the single dimensionless parameter appearing in the Landau-Zener formula (\ref{eq:LZprobability}) and in Eq.\ (\ref{eq:schrod}).  The second-order-in-time dynamics depend additionally on the value of $\Delta / \hbar \omega_{\mathrm{av}}$; panel \textbf{a} has $\Delta / \hbar \omega_{\mathrm{av}}=1/100$, panel \textbf{b} has $\Delta / \hbar \omega_{\mathrm{av}}=1/500$ and panel \textbf{c} has $\Delta / \hbar \omega_{\mathrm{av}}=1/1000$.  Four curves are shown in each panel: the two curves (black and red) that begin on the upper left at $\tau=-25$ both give $A_{L}$. The black curve (which climbs higher) is calculated from Eq.\ (\ref{eq:maxwell3}). The red curve is given by Eq.\  (\ref{eq:schrod}). The other two curves (blue and green) that end on the upper right at $\tau=50$ give $A_{R}$. The blue curve (which is higher) is calculated from Eq.\ (\ref{eq:maxwell3}) and the green curve is from Eq.\  (\ref{eq:schrod}).}
\label{fig:LZcompare}
\end{figure}

\begin{figure}
\includegraphics[width=1.0\columnwidth]{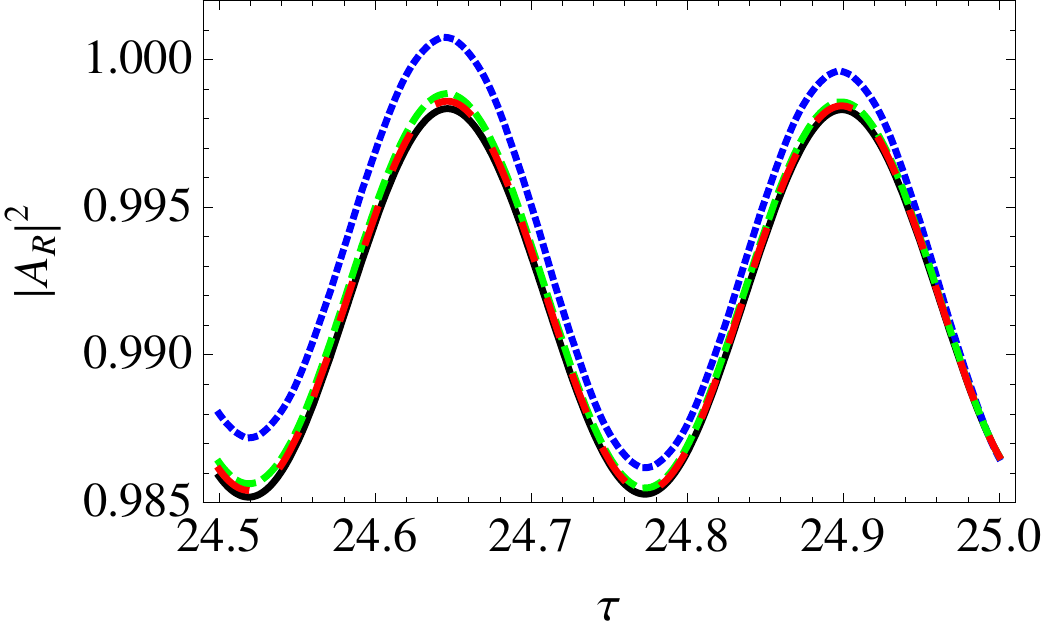}
\caption{(Color online) A zoom-in of Fig.\ \ref{fig:LZcompare} combining the upper right part of all the panels in order to show the time-dependence of $A_{R}$ at the end of the sweep. Like Fig.\ \ref{fig:LZcompare}, this figure illustrates how the solutions of the (second order) Maxwell equation reduce to those of the (first order) Schr\"{o}dinger equation as $\Delta/(\hbar \omega_{\mathrm{av}}) \rightarrow 0$. The solid black curve is calculated from the first order equations (\ref{eq:schrod}), and the remaining curves from the second order equations (\ref{eq:maxwell3}); $\Delta/(\hbar \omega_{\mathrm{av}})=1/100$ (blue short-dashed), $\Delta/(\hbar \omega_{\mathrm{av}})=1/500$ (green medium-dashed), and $\Delta/(\hbar \omega_{\mathrm{av}})=1/1000$ (red long-dashed).}
\label{fig:LZcomparezoom}
\end{figure}

Let us examine the accuracy of the approximations made in obtaining the Schr\"{o}dinger-like equations. Consider a cavity of length $L=100 \, \mu$m and $\alpha/L=0.01$. For the optical wavelength $780$\,nm  we find $n = 128$ and Eq.\  (\ref{Delta approx}) gives $\Delta /\hbar \approx 7 \times10^{11}\mathrm{s}^{-1}$. Furthermore, $E = \Delta$ at the avoided crossing and remains of this order of magnitude. One therefore finds from expansion (\ref{eq:betaexpansion}) that the first correction to the result $\beta_{L/R}=\omega_{\mathrm{av}}$ is of magnitude $ 3 \times10^{-4} \omega_{\mathrm{av}}$ and so relatively small. Also, from expansion (\ref{eq:betadifferenceexpansion}) we see that the first correction to the result $\beta_{R}-\beta_{L}=2 E$ is of second order and equals $5 \times 10^{-8} E$. The smallness of this latter correction is significant because the combination $\beta_{R}-\beta_{L}$ appears in the phases which are generally more sensitive to approximations than amplitudes. 

A further very important check is to see if the neglect of second order derivatives in Eq.\ (\ref{eq:ddotA}) is consistent with our final result (\ref{eq:schrod}). Differentiating Eq.\ (\ref{eq:schrod}), we find that the largest term neglected, namely $\ddot{\tilde{A}}_{L/R}$, has a magnitude of order $(\Delta/\hbar)^2 \tilde{A}_{L/R} $, whereas the smallest term retained, namely $2 i \beta_{L/R} \dot{\tilde{A}}_{L/R}$, has a magnitude of order $(\Delta/\hbar) \omega_{\mathrm{av}}  \tilde{A}_{L/R} $ and so is a factor of $10^4$ bigger. To check this point more systematically, we have numerically solved the first and second order equations and their solutions are compared in Figures \ref{fig:LZcompare} and \ref{fig:LZcomparezoom}. The initial conditions were taken to be $A_{L}=1$ and $A_{R}=0$ at the initial time $\tau=-25$ and the equations were integrated up to the final time  $\tau=25$. Note that in order to solve the second order equations we also require a second set of boundary conditions, e.g.\ the values of the first derivatives of the amplitudes at the initial time. Because the values of these derivatives are required only at the initial time, i.e. before the mirror has started moving, we can take their values from the known solutions for a static mirror. We find that the first derivatives of the amplitudes for a static mirror are given by
\begin{equation}
i\left(\begin{array}{c}\dot{A}_{L}  \\ \dot{A}_{R}\end{array}\right)=\left(\begin{array}{cc} \omega_{\mathrm{av}}-\frac{E}{\hbar} & \frac{\Delta}{\hbar} \\  \frac{\Delta}{\hbar} & \omega_{\mathrm{av}}+\frac{E}{\hbar}  \end{array}\right)\left(\begin{array}{c} A_{L}  \\ A_{R}\end{array}\right) \, .
\label{eq:secondbc}
\end{equation}
This equation should not be confused with Eq.\ (\ref{eq:schrod}).

The first order dynamics generated by the Schr\"{o}dinger-like Eq.\ (\ref{eq:schrod}) depend only upon the value of the combination $\tilde{\vartheta}$. They do not depend upon the value of $\Delta$ by itself. By contrast, the second order dynamics generated by Eq.\ (\ref{eq:maxwell3}) depend upon the separate values of both $\tilde{\vartheta}$ and $\Delta$
(or more precisely $\Delta/\hbar \omega_{\mathrm{av}}$) as can be seen if the time variable in Eq.\ (\ref{eq:maxwell3}) is scaled by $\hbar / \Delta$, like in Eq.\ (\ref{eq:schrod}). Therefore, we expect that as $\Delta/\hbar \omega_{\mathrm{av}}\rightarrow 0$, but $\tilde{\vartheta}$ is held constant, the second order dynamics will reduce to the first order dynamics. This is exactly what happens, as illustrated by the successive panels of Figure \ref{fig:LZcompare}. In  these panels the first order dynamics is held constant by fixing $\tilde{\vartheta}=1$, but $\Delta/\hbar \omega_{\mathrm{av}}$ takes the successive values $1/100,1/500$ and $1/1000$. 

The reduction of the second order Maxwell wave equations to a first order Schr\"{o}dinger-like equation has some broad implications. In the Maxwell theory the amplitudes $A_{L}$ and $A_{R}$ give the electromagnetic energy in the double cavity
\begin{eqnarray}
H_{\mathrm{Maxwell}} & = & \frac{1}{2}\int \varepsilon(x) \vert \mathcal{E}(x,t) \vert^{2} \mathrm{d} t= \frac{\varepsilon_{0}}{2}\left( \vert B_{e} \vert^2 + \vert B_{o}\vert^2 \right) \nonumber \\
& = & \frac{\varepsilon_{0}}{2}\left( \vert A_{L} \vert^2 + \vert A_{R}\vert^2 \right) \, . \label{eq:energy}
\end{eqnarray} 
This energy will not in general be conserved under time-dependent boundary conditions. In quantum mechanics $A_{L}$ and $A_{R}$ have a radically different interpretation in terms of probability amplitudes and do not by themselves give the energy. More discussion of this point is given in Appendix \ref{sec:appendix1}. Of course, the energy will also not in general be conserved in time-dependent quantum mechanics, however the hamiltonian structure of the Schr\"{o}dinger equation means that the quantity $\vert A_{L} \vert^2 +\vert A_{R} \vert^2 =1$ is conserved even for time-dependent problems. This implies that by reducing the Maxwell wave equation to first order in time we artificially enforce energy conservation in our system. 

\begin{figure}
\includegraphics[width=1.0\columnwidth]{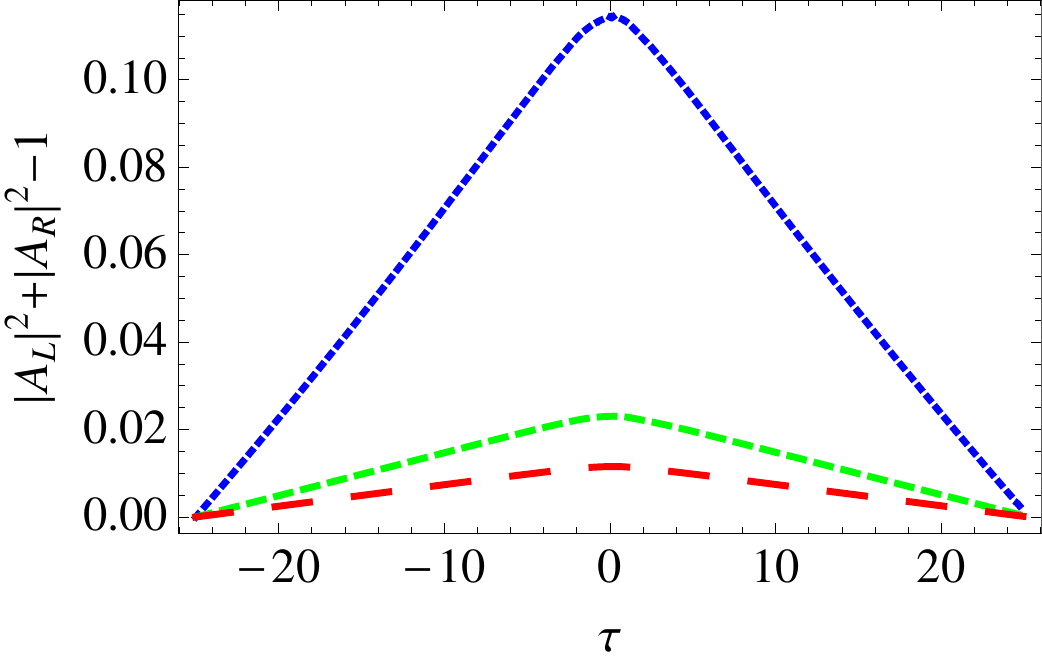}
\caption{(Color online) A plot showing the deviation of $\vert A_{L} \vert^2+\vert A_{R} \vert^2$ from unity as a function of time, as calculated from the full two-mode Maxwell equations (\ref{eq:maxwell3}) which include the second-order time derivatives. As shown in Eq.\ (\ref{eq:energy}), this quantity is proportional to the energy stored in the electromagnetic field. Each curve in this plot has  $\tilde{\vartheta}=1$, but the upper curve (blue, short dashed) has $\Delta/\hbar \omega_{\mathrm{av}}=1/100$, the middle curve (green, medium dashed) has $\Delta/\hbar \omega_{\mathrm{av}}=1/500$, and the lower curve (red, long dashed) has $\Delta/\hbar \omega_{\mathrm{av}}=1/1000$.}
\label{fig:energy}
\end{figure}

Interestingly, whatever the value of $\Delta/\hbar \omega_{\mathrm{av}}$, the agreement between first and second order in time dynamics always seems to be very good at the end of the evolution, as can be seen in Figure \ref{fig:LZcomparezoom}. This is probably connected with the quite precise vanishing of the energy difference between the initial and final states if they are symmetric about the avoided crossing, as illustrated in Figure \ref{fig:energy}. This figure shows how energy is initially pumped into the electromagnetic field from the moving mirror as the avoided crossing is approached and is then removed after it has been passed. Indeed, Figure \ref{fig:energy} also helps us interpret the behaviour of the amplitudes seen in Figure \ref{fig:LZcomparezoom} whereby $\vert A_{L}\vert^2$ and $\vert A_{R}\vert^2$ as calculated from the second order equations climb above the values calculated from the first order equations (which strictly conserve $\vert A_{L}\vert^2+\vert A_{R}\vert^2=1$). This is a good place to remind the reader that the curves plotted in Figures \ref{fig:avoidedcrossingnet} and \ref{fig:avoidedcrossing} showing the avoided crossings of the wave numbers (or equivalently, frequencies) do not correspond to energies as they would in the quantum mechanical case. Indeed, starting with $A_{L}=1$ and $A_{R}=0$ at $\tau=-25$ corresponds to starting approximately in the even global mode on the left of the avoided crossing, and providing the mirror motion is slow, as here, remaining in that mode. The wave number of the even global mode is given by the upper curve in Figure \ref{fig:avoidedcrossing} which \emph{decreases} as it approaches the avoided crossing and \emph{increases} once it has passed it. This is opposite to the behaviour of the energy shown in Figure \ref{fig:energy}, and underscores the difference between the quantum mechanical and classical optical cases.

\section{Quantization}
\label{sec:quantization}

Everything we have discussed in this paper so far has concerned classical electrodynamics. We now wish to briefly see how the hamiltonian is quantized in our formalism. However, as emphasized in Section \ref{sec:frommaxwelltoschrodinger}, we caution the reader again that this hamiltonian approach only strictly applies for the static mirror case because in the dynamic mirror case Maxwell's wave equation provides corrections to the Landau-Zener hamiltonian.

We begin by quantizing in the global basis. The even and odd modes provide the normal modes of the double cavity and so the hamiltonian can be written in the diagonal form
\begin{equation}
H=\hbar \omega_{e}b_{e}^{\dagger}b_{e}+\hbar \omega_{o}b_{o}^{\dagger}b_{o}
\label{Quantum Hamiltonian}
\end{equation}
where $b_{e}$ is the annihilation operator for a photon in the even mode and $b_{o}$ is the annihilation operator for a photon in the odd mode. In contrast to the classical case, the energy now depends upon the frequency of the light. Referring to Appendix \ref{sec:appendix3}, we see that we can transform to the diabatic basis by letting
\begin{eqnarray}
b_{e} & = & \cos \theta \ a_{R}-\sin \theta \ a_{L} \label{eq:photontransforme} \\
b_{o} & = & \sin \theta \ a_{R}+\cos \theta \ a_{L} \label{eq:photontransformo} 
\end{eqnarray}
where $a_{R}$ and $a_{L}$ are the right cavity and left cavity annihilation operators, respectively. We therefore find that
\begin{equation}
\begin{split}
\omega_{e}b_{e}^{\dagger}b_{e} +\omega_{o}b_{o}^{\dagger}b_{o} & =  (\omega_{e}\cos^{2}\theta+\omega_{o}\sin^{2}\theta)a_{R}^{\dagger}a_{R}\\
& +(\omega_{e}\sin^{2}\theta+\omega_{o}\cos^{2}\theta)a_{L}^{\dagger}a_{L}\\ &+\cos\theta\sin\theta(\omega_{o}-\omega_{e})a_{R}^{\dagger}a_{L}\\
& +\cos\theta\sin\theta(\omega_{o}-\omega_{e})a_{L}^{\dagger}a_{R} \, .
\end{split}
\end{equation}
Using the explicit expressions for $\cos \theta$ and $\sin \theta$ given by Eqns (\ref{cos}) and (\ref{sin}), as well as those for $\omega_{e}$ and $\omega_{o}$ given by Eqns (\ref{omega-even}) and (\ref{omega-odd}), we arrive at the second quantized hamiltonian in the diabatic basis
\begin{equation}
H=(\hbar \omega_{\mathrm{av}}+E)a_{R}^{\dagger}a_{R}+(\hbar \omega_{\mathrm{av}}-E)a_{L}^{\dagger}a_{L}+\Delta(a_{R}^{\dagger}a_{L}+a_{L}^{\dagger}a_{R})
\label{Quantum Hamiltonian 4}
\end{equation}
where $\pm E$ are the diabatic energies and $\Delta$ the gap appearing in the Landau-Zener hamiltonian (\ref{LZ-Hamiltonian}).

Writing the hamiltonian in this second quantized form makes it clear that $\Delta$ is the matrix element of $H$ between the left and right diabatic states, i.e.\ $\Delta= \langle \phi_{L} \vert H \vert \phi_{R} \rangle$. When $\Delta$ is small, i.e.\ when the two cavities are weakly coupled,  the transition probability per unit time between the \emph{diabatic} states should obey Fermi's golden rule 
\begin{equation}
\frac{\mathrm{d} P}{\mathrm{d} t}= \frac{2 \pi}{\hbar} \Delta^2 \rho
\label{eq:fermigoldenrule}
\end{equation}
where $\rho=L/2 \pi \hbar c $ is the density of states in either of the uncoupled cavities. This is indeed the case, as can be seen by first noting that in the weak coupling limit we have, from Eq.\ (\ref{Delta approx}), that
\begin{equation}
\Delta \approx \frac{\hbar c}{n \pi \alpha} \approx \frac{2}{k \alpha} \frac{\hbar c}{L} \, ,
\label{eq:1storderDelta}
\end{equation}
where to obtain the second equality we have put $k = 2 \pi n/L $, which is exact in the uncoupled limit. Second, we observe from Eq.\ (\ref{eq:deltatransmission}) that in the weak coupling regime $k \alpha$ is large, and the transmission probability can be written
\begin{equation}
T \approx \frac{4}{k^2 \alpha^2} =  \Delta^2 \left(\frac{L}{\hbar c}\right)^2= \left(  2 \pi \rho \Delta  \right)^2 \, ,
\label{eq:TvsDelta}
\end{equation}
where we have used Eq.\ (\ref{eq:1storderDelta}) to obtain the second equality.
Finally, using the fact that $\mathrm{d} P/ \mathrm{d} t = T / (L/c)$, where $L/c$ is the ``return time'' of a photon in one of the cavities, i.e. the time between collisions with the common mirror, we recover Fermi's golden rule Eq.\ (\ref{eq:fermigoldenrule}).

We saw in Section \ref{sec:frommaxwelltoschrodinger} that in the 
weak coupling regime, i.e.\ when $\Delta/ \hbar \omega_{\mathrm{av}} \rightarrow 0$, the Maxwell wave equation reduces to a Schr\"{o}dinger-like equation for the classical electromagnetic field. The above quantization procedure shows us that in this regime the Schr\"{o}dinger-like equation can be promoted to a true Schr\"{o}dinger equation.

\section{Regimes of validity}
\label{sec:regimes}

According to the Landau-Zener theory, the adiabatic regime occurs when, from Eq.\ (\ref{eq:LZprobability}),
\begin{equation}
\frac{2 \pi \Delta^2}{\hbar \vartheta} \gg 1 \, .
\end{equation}
Various physical interpretations can be given to this condition, as we now show. The rate of change of the separation of the diabatic energies, given in Eq.\ (\ref{eq:thetaapprox}), can be written 
\begin{equation}
\vartheta \approx 2 \frac{\hbar \delta \omega_{\mathrm{dop}}}{L/c}=4 \frac{\hbar  \omega_{\mathrm{av}}}{L/c} \frac{v}{c}
\end{equation}
where $\delta \omega_{\mathrm{dop}} \equiv 2 k v \approx 2 \omega_{\mathrm{av}} v/c $ is the change in angular frequency of a localized mode during the photon return time $L/c$, or, equivalently, the Doppler shift acquired when a light wave reflects from a mirror moving at speed $v$. Then, using Fermi's golden rule Eq.\ (\ref{eq:fermigoldenrule}) to replace $\Delta^2$ with $\mathrm{d}P/\mathrm{d} t$, we find that the adiabatic condition becomes
\begin{equation}
\frac{2 \pi \Delta^2}{\hbar \vartheta} \approx \frac{\pi}{ \delta \omega_{\mathrm{dop}}}  \frac{\mathrm{d} P}{\mathrm{d} t}  \gg 1 \, .
\end{equation}
This says that the change in frequency of the diabatic modes during a photon return time must be much smaller than the transition rate between the diabatic states (roughly speaking, the rate of photon jumps across the common mirror). 

The adiabatic condition must also have a classical interpretation, because Eq.\ (\ref{eq:LZprobability}) applies to classical electromagnetic waves. To find this we use Eq.\ (\ref{eq:TvsDelta}) to replace $\Delta^2$ by the transmission probability $T$ to obtain
\begin{equation}
\frac{2 \pi \Delta^2}{\hbar \vartheta} \approx  \frac{T}{2} \frac{\omega_{\mathrm{FSR}}}{\delta \omega_{\mathrm{dop}}} =  \frac{T}{4} \frac{\omega_{\mathrm{FSR}}}{ \omega_{\mathrm{av}}} \frac{c}{v} \gg 1 \, .  
\label{eq:adiabaticphysical}
\end{equation}
This says that the ratio of the shift in frequency during the photon return time to the free spectral range $\omega_{\mathrm{FSR}} \equiv 2 \pi c /L $ (angular frequency difference between cavity modes) must be much smaller than the fraction of the incident intensity which is transmitted by the common mirror.

\begin{figure}
\includegraphics[width=1.0\columnwidth]{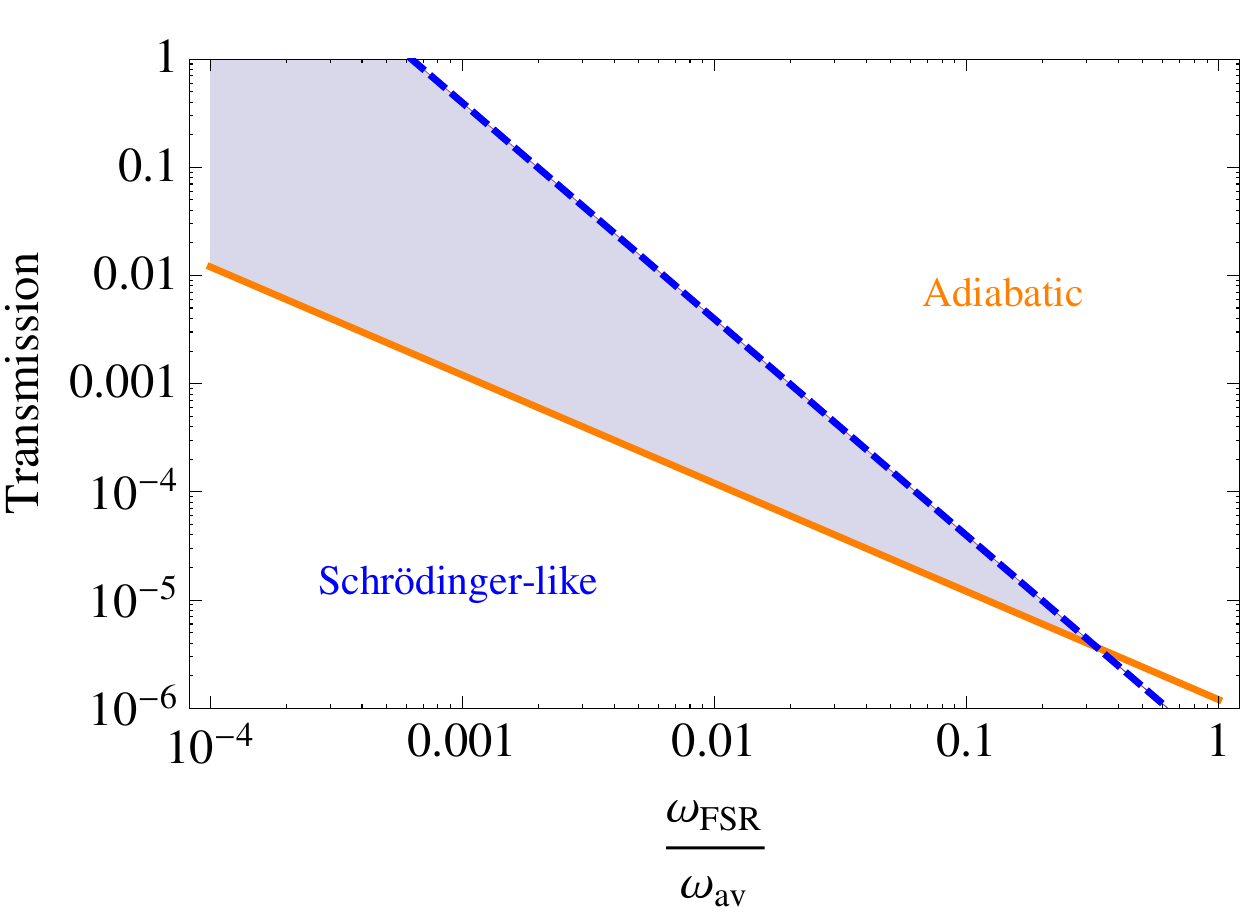}
\caption{A plot showing the various regimes discussed in Section \ref{sec:regimes} as a function of the transmission $T$ of the common mirror and ratio of the free spectral range $\omega_{\mathrm{FSR}}$ to the average optical frequency $\omega_{\mathrm{av}}$. The Schr\"{o}dinger-like equation (\ref{eq:schrod}) is valid below the dashed line, which is given by Eq.\ (\ref{eq:reductioncondition}), the left hand side of which  we have assumed to be equal to $10^4$. This is a very conservative estimate, but is still a factor of 10 less than the value given by assuming the parameters: $L=100 \, \mu$m, $\alpha=0.3 L$, and $\lambda=780$ nm. If the Schr\"{o}dinger-like theory holds, then we can apply the Landau-Zener criterion for adiabaticity, given by Eq.\ (\ref{eq:adiabaticphysical}), and which is plotted as the solid line. To plot this line we assumed adiabaticity sets in when the left hand side is greater than 10 (the exponential dependence of $P_{\mathrm{LZ}}$ certainly ensures that this is the case). The regime of adiabatic evolution of the electromagnetic field is above the solid line. The shaded region shows the regime where the Schr\"{o}dinger-like theory applies and the evolution it predicts is adiabatic. }
 \label{fig:regime}
\end{figure}

In terms of the same physical parameters as introduced above, the condition under which the Maxwell wave equation reduces to the Schr\"{o}dinger-like equation can be written
\begin{equation}
\frac{\hbar \omega_{\mathrm{av}}}{\Delta} \approx \frac{2 \pi}{\sqrt{T}} \frac{\omega_{\mathrm{av}}}{\omega_{\mathrm{FSR}}} \gg 1 \, .
\label{eq:reductioncondition}
\end{equation}
This is a distinctly different condition from Eq.\ (\ref{eq:adiabaticphysical}) for adiabaticity, and this means that it is possible to be either adiabatic or non-adiabatic, but still be in the Schr\"{o}dinger regime where Eq.\ (\ref{eq:schrod}) applies, see Fig.\ \ref{fig:regime}.  

It is perhaps counter-intuitive that Eq.\ (\ref{eq:reductioncondition}) for the validity of the reduction of the Maxwell to the Schr\"{o}dinger-like theory does not depend on the mirror speed. This is because  in this paper  we have chosen to work in the diabatic basis, which is not the eigenbasis, except far from the avoided crossings. When the dynamics of the electromagnetic field is seen from the point of view of the diabatic basis it is, therefore,  not generally stationary, even in the case of a stationary mirror. The characteristic frequency $\Delta/\hbar$ for the intrinsic dynamics in the diabatic basis is set by the coupling between the two cavities,   and this can dominate the extra time-dependence introduced by a moving mirror if the mirror is moving slowly enough. Therefore, for a slowly moving mirror the limiting factor that sets the condition (\ref{eq:reductioncondition}) for the validity of the Schr\"{o}dinger-like theory does not involve $v$.

A direct dependence on the mirror speed only enters the dynamics of diabatic mode amplitudes  in the third term on the right hand side of Eq.\ (\ref{eq:ddotA}), which we have so far ignored.        
Let us estimate the conditions under which we can ignore it. We begin by recalling that condition (\ref{eq:reductioncondition}) comes from comparing the orders of magnitude of the first and second terms on the right hand side of Eq.\ (\ref{eq:ddotA}) for $\ddot{A}$, namely $\ddot{\tilde{A}} \approx O[(\Delta/\hbar)^2 \tilde{A}]$ and $\beta \dot{\tilde{A}} \approx O[(\Delta/\hbar) \omega_{\mathrm{av}} \tilde{A}]$. The magnitude of the third term on the right hand side of Eq.\ (\ref{eq:ddotA}) is 
\begin{equation}
\dot{\beta} \tilde{A} \approx \frac{1}{\hbar} \frac{\mathrm{d} E}{\mathrm{d} t} \tilde{A} =\frac{\vartheta}{2 \hbar} \tilde{A} = \frac{\delta \omega_{\mathrm{dop}}}{L/c} \tilde{A} \, .
\end{equation}
The condition under which this term can be neglected in comparison to the first order time derivative term, is
 \begin{equation}
 O [\beta \dot{\tilde{A}}/\dot{\beta} \tilde{A}] = \frac{\omega_{\mathrm{av}}}{\delta \omega_{\mathrm{dop}}} \frac{\Delta}{\hbar}\frac{L}{c} \approx  \frac{\omega_{\mathrm{av}}}{\delta \omega_{\mathrm{dop}}} \sqrt{T} = \frac{1}{2} \frac{\sqrt{T}}{v/c} \gg 1 \, .
\end{equation}
This condition is obeyed for most practical situations, but does require a certain finite value of the transmission $T$ and/or a non-relativistic mirror. If we assume a mirror with $\alpha/L =0.3$ at a wavelength of 780 nm, then according to Eq.\ (\ref{eq:deltatransmission}) we have $T= 7 \times 10^{-5}$, implying that the Schr\"{o}dinger-like equation (\ref{eq:schrod}) holds if $v/c  \ll 0.008$.

\section{Finite Width Mirror}
\label{sec:FiniteWidthMirror}

\begin{figure}
\includegraphics[width=1.0\columnwidth]{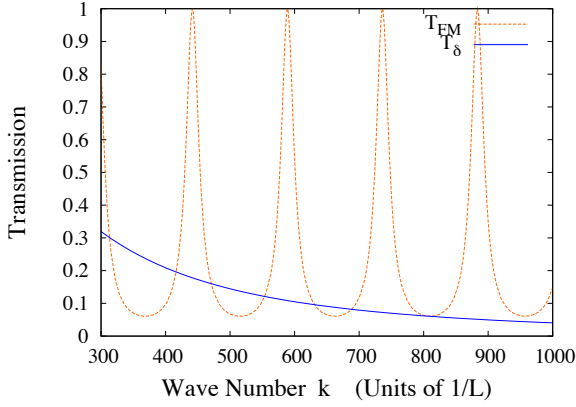}
\caption{A comparison of the transmission probability $T$ for the finite and $\delta$-function mirrors, as given by Eqns (\ref{eq:finitetransmission}) and (\ref{eq:deltatransmission}), respectively.  The resonances of the finite mirror occur at the peaks of its transmission function. For the parameters used in this plot, which are the same as those given in the caption of Fig.\ \ref{fig:DFB}, the two curves intersect near a minimum of the finite barrier transmission. This means that the two transmission probabilities remain in good agreement for a larger range of $k$ than they might otherwise do.}
 \label{fig:transmission}
\end{figure}

We now go beyond the $\delta$-function model introduced in Section \ref{sec:deltafunctionmirrormodel} and consider the case of a mirror with a finite width. For many purposes the $\delta$-function model is perfectly sufficient and has the advantage of being mathematically simple, but there are phenomena which it is not capable of describing such as resonances within the mirror. In this section we compare the wave numbers of the normal modes produced with the $\delta$-function model  with those of a model of a mirror as a uniform slab of dielectric material of finite width (we have in mind something like the SiN membrane used in \cite{thompson08}, although perhaps somewhat thicker). Our main purpose is to check how well the $\delta$-function model reproduces the predictions of the finite width model. Of course, the finite width model considered here does not accurately capture the structure of an actual piece of dielectric material, but it does take the next step in that direction.

As a model for a mirror of finite width we take the same basic model as considered in Section \ref{sec:deltafunctionmirrormodel} except that we replace the $\delta$-function at $x=0$ by a uniform slab of dielectric of width $2M$ centred at $x=0$. The corresponding permittivity function is given by 
\[\varepsilon(x)=\begin{cases}
\varepsilon_{0}\left(1+h\right) & -M \leq x \leq M \\
\varepsilon_{0} & -L_{1} < x < -M\ \mbox{and} \ M < x < L_{2}\\
\infty & \mbox{elsewhere} \end{cases}\]
where $n_{r}=\sqrt{1+h}$ is the refractive index inside the slab.
Imposing the boundary conditions $U_{n}(-L_{1})=U_{n}(L_{2})=0$, the
normal modes can be written as:\[
U_{n}(x)=\begin{cases}
A \sin[k(x+L_{1})] & -L_{1} \leq x \leq -M\\
B \sin[n_{r}kx+\phi_{1}] & -M \leq x \leq M\\
C \sin[k(x-L_{2})] & M \leq x \leq L_{2} \, .
\end{cases}\]
The parallel component of the electric field is continuous at the mirror surface, so that provides us with two more boundary conditions upon the mode function, namely continuity of $U_{n}(x)$ at $x=\pm M$. Another two conditions are provided by the continuity of the parallel component of the magnetic field at the mirror surfaces (assuming the mirror is non-magnetic so that $\mu=\mu_{0}$) which, when combined with Faraday's law $\nabla \times \mathbf{\mathcal{E}}= - \partial \mathbf{B} / \partial t$, lead to the continuity of the first derivatives of $U_{n}(x)$ at $x=\pm M$. Implementing these boundary conditions we
find that the allowed wave numbers $k$
 satisfy the following equation 
 \begin{equation}
\begin{split}
-n_{r}^{2} & \sin(2n_{r} k M) \sin\left[k\left(\frac{L}{2}-M+\frac{\Delta L}{2}\right)\right]\times
\\ &
\sin\left[k\left(\frac{L}{2}-M-\frac{\Delta L}{2}\right)\right] \\ & +n_{r} \cos(2n_{r}k M)\sin[k(L-2 M)]\\ & +
\sin(2n_{r} k M)\cos\left[k\left(\frac{L}{2}-M+\frac{\Delta L}{2}\right)\right]\times
\\ & 
\cos\left[k\left(\frac{L}{2}-M-\frac{\Delta L}{2}\right)\right]=0 \, ,
\end{split}
\end{equation}
which is the equivalent of Eq.\ (\ref{transcendental2}), but for the finite mirror case.
The mirror resonances occur when $2M k n_{r} = l \pi$, where $l=1,2,3, \ldots$, and so the distance in $k$-space between mirror resonances is $\pi/(2Mn_{r})$. This is much larger than the distance between cavity resonances $\pi / L$ providing $L \gg 2 M n_{r}$.

The transmission $T=|t|^2$ of the finite mirror, given by Eq.\ (\ref{eq:finitetransmission}) in Appendix \ref{sec:appendix2} as $\left(1+\frac{(n_r^2-1)^2}{4 n_r^2}\sin^2(2n_r k M)\right)^{-1}$, is shown as a function of wave number by the dotted line in Fig.\ \ref{fig:transmission}. These are the usual transmission fringes of a Fabry-Perot resonator. By contrast, the power transmission of the $\delta$-function mirror, given by Eq.\ (\ref{eq:deltatransmission}) as $\left(1+\alpha^2 k^2/4\right)^{-1}$, is the monotonic function of wave number illustrated by the solid line in Fig.\ \ref{fig:transmission}. The two have the same transmission when
\begin{equation}
k^2\alpha^2=\frac{(n_r^2-1)^2}{n_r^2}\sin^2(2n_r k M).
\end{equation}
In general, this equality is satisfied only over narrow wave number regions where the two curves cross, but in Fig.\ \ref{fig:transmission} we have chosen a value of $\alpha$ such that they have similar transmissions over a wider range of wave numbers in the vicinity of $k=800$. In this region, therefore, the $\delta$-function mirror can be a simple model for the more complicated finite mirror case. Of course, the phase shifts produced by the two mirrors are not equal, as one can see from Eqns (\ref{eq:finitetransmission}) and (\ref{eq:deltatransmission}), and indeed the finite mirror phase changes rapidly in the vicinity of the Fabry-Perot transmission peaks. At the transmission minima, however, the rate of change of the phase shift with $k$ is much slower. Therefore, over the range of wave numbers near $k=800$ where they have roughly equal values of $T$, the two mirrors produce phase shifts that do not change appreciably and are equivalent to small fixed displacements of the mirror positions.

\begin{figure}
\includegraphics[width=1.0\columnwidth]{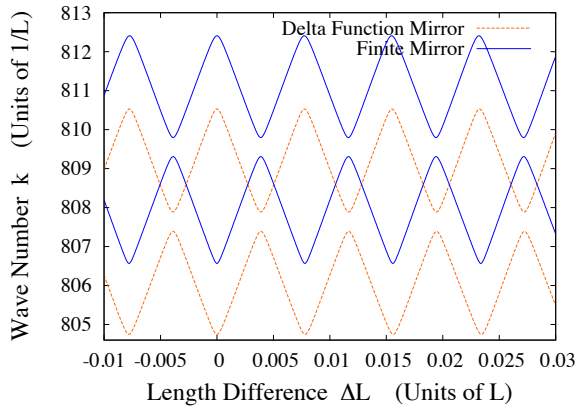}
\caption{(Color online) Comparison between the wave number structure produced by a mirror of width $2M$ (solid blue curves) and
the $\delta$-mirror (dashed red curves) inside a double cavity. This plot is for the case when we are far from a mirror resonance. In order to make the comparison we matched the transmission probabilities $T$ for each mirror over the relevant range of wave numbers as best we could (see Fig.\ \ref{fig:transmission}). Up to a relative vertical shift, the agreement is very good, including  the magnitude of the gap $2 \Delta$ at the avoided crossings. In the calculation for the $\delta$-function mirror we set  $\alpha/(L-2M) =0.009735$. In the finite mirror calculation we set $M/(L-2M)=0.001334 $ and $2 n_{r}^2 M/(L-2M)= 0.1707$.  Note that to make this plot clearer we have only included the wave numbers for a single pair of modes in each case.}
\label{fig:DFB}
\end{figure}

Fig.\ \ref{fig:DFB} shows the wave number curves for these two cases. Since we are comparing the transmission function of a $\delta$-mirror to that of a mirror of width $2M$, we have reduced the total cavity length for the $\delta$-mirror case so that $L \rightarrow L-2M$. This means that the optical path length outside the mirror is the same in both cases. Nevertheless, the different phase shifts imposed upon the light when it traverses the two different types of mirror produces a small vertical shift in the wave numbers.
The two sets of curves are otherwise almost identical, including the gap $2\Delta$ at the avoided crossings.

\begin{figure}
\includegraphics[width=1.0\columnwidth]{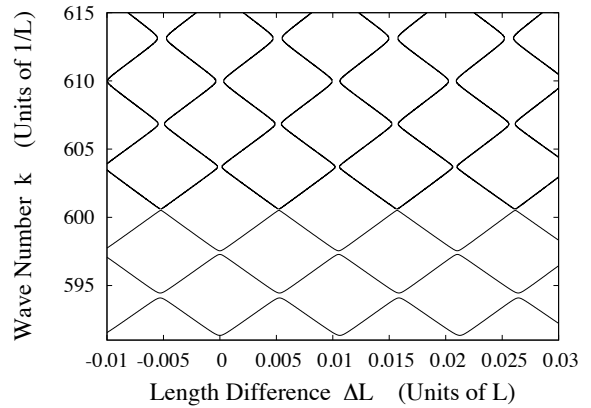}
\caption{The wave number structure produced by a mirror of width $2M$ inside a double cavity. This plot includes wave numbers for which there is a mirror resonance. At the bottom we see the familiar avoided crossing structure which couples even and odd modes with the same index $n$, but for higher wave numbers we pass into resonance and the structure changes to one where there are vertical avoided crossings that allow for the possibility of adiabatically coupling localized cavity modes with different values of the index $m$. The parameters used to make this plot were M/(L-2M)=0.0005005, and $2 n_{r}^2 M/(L-2M)=0.01001$.}
\label{fig:OnResonance}
\end{figure}

A very different wave number structure occurs in the resonant case when an integer multiple number of half wavelengths fit inside the mirror, as shown in Fig.\ \ref{fig:OnResonance}. At the bottom of the figure we have the familiar horizontal avoided crossings where an even and an odd mode with the same value of the index $n$ are coupled. However, as we increase the wave number the light approaches resonance with the mirror and we find that the amplitude $B$ of the light inside the mirror becomes large.  Intriguingly, we see that the avoided crossings become vertical, meaning that we can adiabatically transfer between localized modes with different values of $n$.

\section{Experimental feasibility}
\label{sec:experimentalfeasibility}
There are many different ways to realize the double cavity model discussed here in an experiment. The key challenge in all of them is likely to be achieving adiabatic transfer before the light leaks out of the system, e.g.\ through the end mirrors, which hitherto we assumed to be perfect, but which in reality never can be. Remarkably, this challenge has already been overcome in the experiment \cite{preble07} performed in a $6\mu$m diameter ring resonator made out of silicon where ultrafast electro-optical control of the refractive index allows the effective optical length of the cavity to be changed on time scales shorter than 10 ps, which is the lifetime of a photon in their resonator. In this way, the wavelength of light injected into the cavity was adiabatically shrunk by up to 2.5nm before leaking out. The change in refractive index was achieved by shining a laser onto the silicon thereby creating free charge carriers which, via plasma dispersion, strongly modified its optical properties. The same electro-optical effect can also be generated electrically \cite{xu05}.

In order to make a double cavity out of this system we imagine coupling two ring cavities together. It is not necessary to displace a mirror: the difference in cavity lengths $\Delta L=L_{1}-L_{2}$ can be controlled by changing the effective optical lengths of each cavity in a concerted fashion via the electro-optical effect; see Appendix \ref{sec:appendix2b}.  The degree to which the light can be localized on one side of the double cavity or the other is set by the dimensionless parameter $\alpha/L$, as discussed in Section \ref{sec:transferratio}. For a central ``mirror'' such that $\alpha/L=0.01$, the maximum amplitude  ratio that can be achieved when $n=128$ is approximately $\vert A/B \vert \approx 10$, and so the probability of finding a photon on one side versus the other is $100:1$. For a double cavity of length $L=100 \mu$m we would then require $\alpha=1 \mu$m, and at a wavelength of $780$nm Eq.\ (\ref{eq:deltareflectivity}) says that this corresponds to a central ``mirror'' with a reflectivity of $R=0.94$. This degree of coupling is readily achieved.

\begin{figure}
\includegraphics[width=1.0\columnwidth]{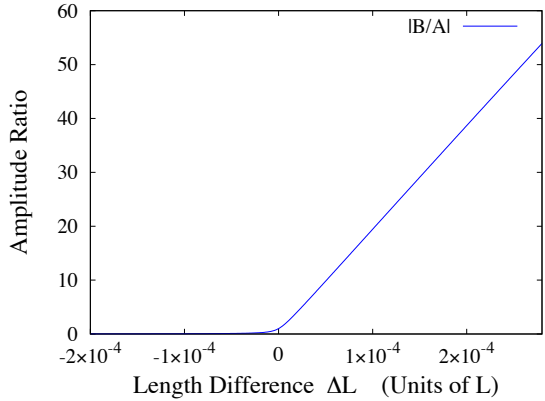}
\caption{A zoomed-in plot of the relative amplitudes versus mirror displacement $\Delta L$.  It is clear that one only needs small displacements to achieve
large amplitude transfers.  Here $\alpha/L=0.3$, and $n=128$.}
\label{fig:B/Azoom}
\end{figure}

There are some advantages to using a  more reflective common mirror. Not only does the maximum  localization become larger, but also a significant degree of localization can be achieved for a much smaller value of $\Delta L$.  This is illustrated in Figure \ref{fig:B/Azoom} for $\alpha/L=0.3$ where we see that an amplitude ratio of 20:1 can be achieved with a displacement of only  $\Delta L= 10^{-4} L$ (the maximum amplitude ratio that can be achieved with $\alpha/L=0.3$ and $n=128$ is 240:1).

\begin{figure}
\includegraphics[width=1.0\columnwidth]{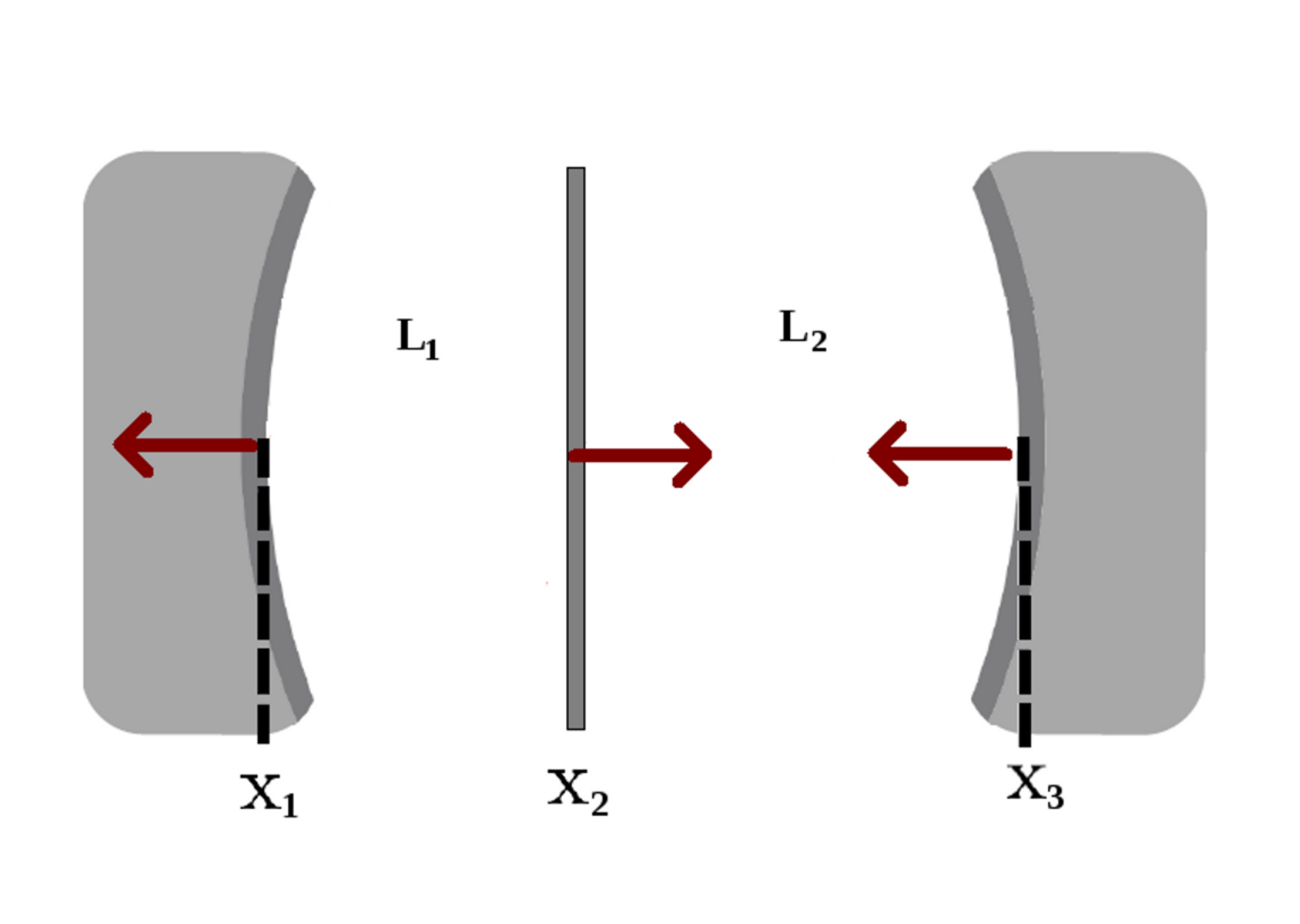}
\caption{(Color online) By simultaneously moving all 3 mirrors, it is possible to double the effective speed of the common mirror, i.e.\ quadruple the rate of change of the relative cavity length $\Delta L(t)=L_{1}(t)-L_{2}(t)$.}
\label{fig:CavityTripleMove}
\end{figure}

Another possible way to realize the double cavity model is to displace the mirrors, e.g.\ by piezo-electric motors. These are routinely used to control mirror positions in order to actively stabilize the lengths of ultrahigh finesse optical cavities against vibrations \cite{ottl06}. 
However, it is an obstacle that only rather low mirror speeds can be achieved. Piezoelectric motors can accurately move small distances (nm), but the current fastest ones of which we are aware only reach speeds of $v=0.8$m/s \cite{PI}. This speed can be effectively doubled if all three mirrors are moved at once as shown in Fig.\ \ref{fig:CavityTripleMove}. On the other hand, the distances the mirrors need to move are only a fraction of a wavelength: in order to transfer light from one side of a double cavity to the other we only need 
the mirror to move by a distance of up to $\lambda/4$, i.e. $\Delta L$ needs to change by $\lambda/2$.
This gives us a transfer time of $t=\lambda/4v \approx 120$ns if we take $v=1.6$\,m/s. 

For the case when $\alpha/L=0.01$ with a cavity of length $L=100 \mu$m, the cavity mode corresponding most closely to light with a wavelength of $780$\,nm has $n = 128$.
Inserting these numbers into Eq.\ (\ref{Delta approx})  gives $\Delta/\hbar \omega_{\mathrm{av}} =3 \times 10^{-4} $. Combining this with Eq.\ (\ref{eq:thetaapprox}), we see that at a velocity of $v=1.6$m/s we have $\hbar \vartheta_{n} / \Delta^2=2 \times 10^{-4}$ and thus the transfer is perfectly adiabatic, i.e.\ the Landau-Zener transition probability $P_{\mathrm{LZ}}$ given by Eq.\ (\ref{eq:LZprobability}) is utterly negligible. Non-adiabatic transitions only become significant in the ultrafast electro-optical case: if we assume the same parameters  as for the Fabry-Perot cavity, except that the transfer takes 10 ps, and thus a rate of change of cavity length of $\mathrm{d} \Delta L / \mathrm{d} t = 39,000$ m/s, we find  $\hbar \vartheta_{n} / \Delta^2=6.8$, and a non-adiabatic transfer probability of $\exp[-2 \pi \Delta^2/(\hbar \vartheta)]=0.4$. In Appendix \ref{sec:appendix5} we make the observation that such a rapidly changing refractive index probably necessitates the use of a modified Maxwell wave equation for its proper theoretical description. 
 
At the relatively slow transfer speeds of piezo-electric motors the leakage of light out of the end mirrors becomes important. To estimate this effect we note that a number of recent experiments have realized Fabry-Perot microcavities with lengths below $200 \mu$m and finesses of up to $\mathcal{F}=10^6$ \cite{brennecke07,colombe07,trupke07,murch08}. The decay rate of the electric field amplitude in a cavity is given by $\kappa= c \pi / 2 d \mathcal{F}$. Being optimistic and assuming that for cavity of length $100 \mu$m we can achieve a finesse $\mathcal{F}=10^6$, one finds $\kappa = 5 \times 10^6$$s^{-1}$, and hence the field decays to only $\exp[-\kappa t]=0.6$ of its original value during a transfer lasting $120$ns. Thinking in terms of transferring a single photon, the photon has an $\approx 60\%$ probability of escaping the cavity by the end the transfer. 

The situation improves when the central mirror is made more reflective, as mentioned above. For $\alpha/L=0.3$, which is still an order of magnitude less reflective than the end mirrors, and with a total change of the cavity length difference of  $\Delta L=0.0002 L$ (i.e.\ moving between the points $\Delta L = \pm 0.0001 L$ either side of $\Delta L=0$) for a cavity of length $L=100 \mu$m, the transfer time is $t= \Delta L /2v= 6$ns when $v=1.6$\,m/s, leading to a $6 \%$ probability that the photon will escape during the transfer.

\section{Discussion and conclusions}
\label{sec:conclusion}

The optical double cavity is a closely related system to the double well potential in quantum mechanics, although, as discussed in Appendix \ref{sec:appendix1}, the common central mirror in the cavity case does not act as a tunneling barrier. The  wave numbers (frequencies) of the global modes inside the double cavity form a net of avoided crossings as a function of the difference in length of the two cavities, and this permits the adiabatic transfer of light from one side of the common mirror to the other. Alternatively, if the transfer is halted right at an avoided crossing the light is left in a superposition of being in both cavities. The double cavity is  arguably, therefore, the simplest possible prototype for a quantum network which can transfer photons on demand from one node to another or can place them into superpositions of being on different nodes.

Using a $\delta$-function model for the common mirror, we have obtained analytic results such as those given in Eqns (\ref{even approx}) and (\ref{odd approx}) for the wave numbers close to an avoided crossing.
An expression for the maximum possible transfer ratio of light is given in Eq.\ (\ref{eq:maxA/Bapprox}), and depends on the transmission properties of the central mirror, as parameterized by the dimensionless quantity $\alpha/L$. Whilst the $\delta$-function model gives an adequate description for many purposes, we saw that when the central mirror is modelled as a dielectric slab of finite thickness there is the possibility of mirror resonances which dramatically change the structure of the net of avoided crossings for the global wave numbers. In the vicinity of a mirror resonance the horizontal avoided crossings become vertical avoided crossings, suggesting a way to adiabatically move up the ladder of cavity resonances.

At first sight, the Schr\"{o}dinger and Maxwell wave equations seem to give incompatible descriptions of the time evolution of the electromagnetic field because they are first- and second-order in time, respectively. The first-order time evolution found in quantum theory is a consequence of its hamiltonian structure, which is deeply connected with the conservation of probability. The standard method for quantizing the Maxwell wave equation $\partial_{x}^2 \mathcal{E}-(n_{r}^2/c^2)\partial_{t}^2 \mathcal{E}=0$, where $n_{r}(x)$ is the refractive index, is to give it hamiltonian structure: a separation of variables of the form $\mathcal{E}(x,t)=\sum_{n} B_{n}(t)U_{n}(x)$  leads to the two equations $\partial_{x}^2 U_{n}+n_{r}^2 U_{n}=0$ and $\partial_{t}^2 B_{n}+\omega_{n}^2 B_{n}=0$, where $\omega_{n}^2=c^2 k_{n}^{2}$ is the separation constant. The first equation gives the normal (global) modes, and the second is the equation of motion for a simple harmonic oscillator. This can be transformed into hamiltonian form by introducing the variables  $q=\sqrt{\hbar/2m\omega}(B^{\star}+B)$ and $p=i\sqrt{m \hbar \omega/2}(B^{\star}-B)$ whose time evolution is given by Hamilton's equations $\dot{x}=\partial_{p}H$ and $\dot{p}=-\partial_{x}H$, which are first order in time. The final step is, of course, to use the commutator $[x,p]=i\hbar$. This prescription is straightforward to apply in cases where the refractive index is independent of time, but when it depends on both time and space the Maxwell wave equation does not separate, there are no normal modes, and a simple harmonic oscillator equation for the field amplitudes is harder to obtain. The time dependent problem can be tackled in  number of ways, e.g.\ by working in a time-independent basis. However, for slow time dependence it can be easier to work in the adiabatically evolving basis  $U_{n}(x,t)$, given by the instantaneous solutions to Maxwell's wave equation at time $t$, so that electric field can be written $\mathcal{E}(x,t)=\sum_{n} c_{n}(t)U_{n}(x,t)\exp[-i \int_{0}^{t} \omega_{n}(t')dt']$.

In this paper our primary concern was not the quantization of the electromagnetic field, but rather the application of a mathematical solution, namely, that of the Landau-Zener problem, to the transfer of classical electromagnetic field modes in a double cavity using a time-dependent mirror (or, equivalently, time-dependent refractive indices, see Appendix \ref{sec:appendix2b}). In order to apply the Landau-Zener model we need a first-order-in-time equation of motion. Our approach, which is closely related to using the adiabatic basis mentioned above, was to use a `paraxial approximation in time' (slowly varying envelope approximation) to reduce the Maxwell wave equation to the mathematical form of the  time-dependent Schr\"{o}dinger equation, given by Eq.\ (\ref{eq:schrod}). This reduction holds in the regime $\Delta/\hbar \omega_{\mathrm{av}} \rightarrow 0$ and assumes that the dimensionless parameters $\tilde{\vartheta} \equiv \hbar \vartheta/ \Delta^2$, $\tau_{i} \equiv \Delta t_{i}/\hbar$ and $\tau_{f} \equiv \Delta t_{f}/\hbar$ are held constant. $\vartheta$ is the quantity which appears in the Landau-Zener formula (\ref{eq:LZprobability}), and $t_{i}$ and $t_{f}$ are the initial and final times, respectively.  $\tilde{\vartheta}$ and $\tau$ appear in both the full Maxwell equation and  the Schr\"{o}dinger-like equation, but $\Delta/\hbar \omega_{\mathrm{av}}$ only appears in the full Maxwell equation. The reduction, therefore, does not affect the (lower order) Schr\"{o}dinger dynamics. This process may be likened to the reduction of quantum mechanics to classical mechanics by letting $\hbar \rightarrow 0$, whilst holding the classical mechanics (i.e.\ the action) fixed \cite{berry+mount}. One consequence of the reduction to the Schr\"{o}dinger-like equation (\ref{eq:schrod}) is that energy is artificially conserved, as discussed in Section \ref{sec:frommaxwelltoschrodinger}. In Section \ref{sec:quantization} we saw that as a by-product of our approach we were able to quantize the hamiltonian in the regime $\Delta/\hbar \omega_{\mathrm{av}} \rightarrow 0$, and therefore promote the Schr\"{o}dinger-like  equation to a true Schr\"{o}dinger equation for the quantum amplitudes of the electromagnetic field in this regime. One of the lessons  our treatment of the time-dependent problem teaches us is that outside of the weak-coupling regime one cannot simply take the solutions from the time-independent Maxwell's wave equation and plug them into the Landau-Zener theory in order to obtain time-dependent properties of the electromagnetic field.

There is a simple physical interpretation of the fact that the reduction from second- to first-order-in-time dynamics holds in the regime $\Delta/\hbar \omega_{\mathrm{av}} \rightarrow 0$. In this regime the gap is  small, which corresponds to weak coupling between the two cavities, as is clear from Eq.\ (\ref{eq:TvsDelta}), which says that the transmission probability  $T \propto \Delta^2$ when $T$ is small. When the transmission is small the time evolution of the amplitudes $A_{L}(t)$ and $A_{R}(t)$ becomes slow,  their higher time derivatives can be neglected, and hence the reduction to first-order-in-time is valid. As pointed out in Section \ref{sec:regimes}, the weak-coupling requirement $\Delta/\hbar \omega_{\mathrm{av}} \rightarrow 0$ for the validity of the Schr\"{o}dinger-like equation is an intrinsic feature of working in the diabatic basis and is not related to the motion of the mirror. The weak-coupling requirement could be eliminated by working in the adiabatic basis, but then the time dependence of the adiabatic basis must be taken into account.

In this paper we have focussed on the optical regime. However, the basic idea of adiabatic transfer of electromagnetic fields between cavities can also be applied at other wavelengths, e.g.\ microwave, which is relevant both to micromaser experiments \cite{raimond01,weidinger99}, as well as to the new field of circuit QED where superconducting qubits play the role of  ``artificial atoms''  in microwave resonators \cite{walraff04}. Micromaser cavities with $Q$ factors as high as $Q=3 \times 10^{10}$ have been demonstrated \cite{weidinger99} and this gives a photon lifetime of $0.2$s, relieving some of the problems relating to photon decay during the adiabatic transfer.

\begin{acknowledgements}
We gratefully acknowledge B. M. Garraway, J. Larson, S. Scheel, J. E. Sipe and A.-C. Shi for discussions. Funding was provided by the Ontario Ministry of Research and Innovation,  and NSERC (Canada); by EPSRC,  and the Royal Society (U.K.); and by the HIP project of the European Commission.
\end{acknowledgements}

\appendix
\section{The connection to the double well problem in quantum mechanics}
\label{sec:appendix1}

In this appendix we compare and contrast classical light waves obeying the Maxwell equations to quantum matter waves obeying the Schr\"{o}dinger equation. These differences are well known, but we include them here for completeness.  In particular, in the main body of this paper we have alluded to the existence of a close connection between the problem of light in a double cavity, and the problem of a quantum particle in a double well potential. In fact, as we shall see in this appendix, the true correspondence is not to a quantum particle tunneling between two classically allowed regions through a classically forbidden potential \emph{barrier}, but rather to a quantum particle passing between two classically allowed regions separated by a potential \emph{well}, all three regions being classically allowed. 

Consider on the one hand the Maxwell wave equation obeyed by plane light waves with vacuum wave number $k= \omega/c$ propagating in one dimension in a medium with refractive index $n_{r}(x)$  
\begin{equation}
\frac{\mathrm{d}^2 \mathcal{E}}{\mathrm{d} x^2}+ k^2n_{r}^{2}(x) \mathcal{E}=0 \ .
\end{equation}
On the other hand, consider the time-independent Schr\"{o}dinger equation for a particle with energy $E=\hbar^{2}k^2 /2m$ and subject to a potential $V(x)$  
\begin{equation}
\frac{\mathrm{d}^2 \psi}{\mathrm{d} x^2}+ k^2 \left(1- \frac{V(x)}{E} \right) \psi=0 .\
\end{equation}
An obvious difference between the two is that in the Maxwell case the eigenvalue $k^2$ multiplies the spatially dependent term $n_{r}(x)$, whereas in the Schr\"odinger equation $k^2$ does not multiply $V(x)$, despite the suggestive way we have written it here. Nevertheless, the two equations are equivalent if we can identify
\begin{equation}
n_{r}^{2}(x)=1-\frac{V(x)}{E} \ .
\end{equation}
This equation forms the basis of matter wave optics. Assuming that the refractive index obeys $n_{r}^2>1$, we see that the analogy between and light waves and matter waves only holds if $V(x)<0$. Thus, the potential ``barrier'' presented by the mirror in the double cavity is, in fact, a potential well if we consider the problem in quantum mechanical terms. This explains why we find the odd adiabatic solution lies below the even adiabatic solution (see Figure \ref{fig:avoidedcrossing}), in contradiction to our experience of the double well problem in quantum mechanics.

In the specific case of the $\delta$-function mirror, which is described by the dielectric function $\varepsilon(x)=\varepsilon_{0} n_{r}^{2}(x)=\varepsilon_{0}(1+ \alpha \delta(x))$, we have
\begin{equation}
\frac{\mathrm{d}^2 \mathcal{E}}{\mathrm{d} x^2}+ k^2(1+\alpha \delta(x)) \mathcal{E}=0
\label{eq:maxwelltimeindependent}
\end{equation}
whilst the equivalent problem in quantum mechanics has a particle of energy $E$ subject to a potential $V(x)=\beta \delta(x)$, giving rise to the
Schr\"{o}dinger equation 
\begin{equation}
\frac{\mathrm{d}^2 \psi}{\mathrm{d} x^2}+ k^2 \left(1- \frac{\beta \delta(x)}{E} \right) \psi=0 .\
\label{eq:schrodingertimeindependent}
\end{equation}
The two equations can be considered as equivalent if $\alpha = - \beta/E $.

Other differences between the Maxwell and Schr\"{o}dinger cases include the normalization conditions and also the interpretation of the wave number as an energy. Starting with the normalization, the wave function $\psi$ in the above equations  obeys $\int_{-L_{1}}^{L_{2}} \vert \psi(x) \vert^2 \mathrm{d} x = 1$, whereas the electric field $\mathcal{E}$ obeys Eq.\ (\ref{normalization}). This difference arises, of course, because $\psi$ is a probability amplitude and probability must be conserved, whereas $\mathcal{E}$ has no such interpretation. Turning now to the wave number $k$, if we were solving a quantum mechanical problem, the curves in Figures \ref{fig:avoidedcrossingnet} and \ref{fig:avoidedcrossing} showing $k$  versus mirror displacement $\Delta L$ would become energy curves. No such connection exists in the Maxwell case. Indeed, in quantum mechanics energy and wave number are intimately connected, e.g.\ in free space $E=\hbar^{2}k^2/2m$, but on the other hand a quantum wave has an energy independent of its amplitude.  The exact converse is true for classical light waves: their energy is completely independent of wave number, but depends rather on the square of their amplitude.
A wavelength dependence of the energy of light waves only enters the story when the light wave is quantized via the photon energy $E=\hbar ck$.

The problem discussed in this paper is purely linear since we have not included atoms in the cavities which can couple to the light field, nor have we considered any nonlinear optical elements. However, if a nonlinearity is introduced into the quantum mechanical double well problem then the resulting system has a close connection to the Josephson junction \cite{smerzi97}. A number of authors have exploited this analogy between coupled optical cavities (or fibres) with nonlinearity and the Josephson junction problem to predict effects such as self trapping of the light in the cavities \cite{larson10,chefles95,gerace09}. If an array of cavities is considered, then the possibility exists to realize a bosonic Hubbard model, for which the self-trapping effect develops into a full-blown bosonic version of the Mott-Insulator quantum phase transition \cite{hartmann06,greentree06}.

\section{Transmission amplitudes for lossless dielectric mirrors}
\label{sec:appendix2}

In this appendix we collect together some results on the transmission  functions of $\delta$-function and finite width mirrors. The problem of light scattering in one dimension from a finite mirror with a homogeneous refractive index is essentially the same as that of quantum particles scattering from a square potential well \cite{schiff} (see also Appendix \ref{sec:appendix1}). For concreteness, consider an incoming wave $A \exp (i  k x)$, a reflected wave $B \exp (-i  k x)$, and a transmitted wave $C \exp (i k x)$. The thickness of the mirror is taken to be $2M$, and it is assumed to be made of a lossless dielectric material with refractive index $n_{r}$. The transmission amplitude $t$ is found to be
\begin{equation}
t \equiv \frac{C}{A} = \frac{\exp (-i 2 M k )\exp (i \arctan[\frac{n_{r}^2+1}{2n_{r}}\tan 2Mk n_{r}])}{\sqrt{1+\frac{(n_{r}^2-1)^2}{4n_{r}^2} \sin^{2} (2 M k n_{r})}} \ .
\label{eq:finitetransmission}
\end{equation}
The transmission probability $T$ for light is given by $T \equiv \vert t \vert^2$, and this is related to the reflection probability $R$ by $R  = 1- T$. 

Performing a similar calculation for a $\delta$-function mirror, as specified by the dielectric function $\varepsilon(x)= \varepsilon_{0} (1+\alpha \delta (x))$, we find
\begin{equation}
t= \frac{1}{1-i k \alpha/2}= \frac{\exp(i \arctan[ k \alpha/2])}{\sqrt{1+k^2 \alpha^2/4}} .\
\label{eq:deltatransmission}
\end{equation}
This latter result can also be obtained from the transmission amplitude for a finite mirror as given above if we let $M \rightarrow 0$, and $n_{r} \rightarrow \infty$, but keep the quantity $n_{r}^2 M$ fixed. This procedure allows us to identify the parameter $\alpha$ as being given by
\begin{equation}
\alpha = 2 M n_{r}^2 \ .
\label{eq:alphathin}
\end{equation}
This only holds providing $2Mk n_{r} \neq m \pi $ where $m=0,1,2\ldots$, i.e. avoiding resonances whereby an integer number of half wavelength fit inside the mirror.

\section{A double cavity made of two coupled waveguides with controllable refractive indices}
\label{sec:appendix2b}

In this appendix we consider a  double cavity formed from two coupled waveguides, each made of a dielectric whose refractive index can be controlled, e.g.\ via the electro-optic effect. The \emph{optical} lengths of the two cavities can be independently controlled by modulating the refractive index in each waveguide, and this allows the optical modes to be transferred between the two waveguides without moving the mirror. Consider the situation where each waveguide has the same length $L/2$, and one has the refractive index $n_{1}$ and the other has refractive index $n_{2}$. The original double cavity model, as depicted in Fig.\  \ref{fig:cavitysetup}, can be converted into the waveguide problem by substituting 
\begin{eqnarray}
L_{1} & \rightarrow  & n_{1} \frac{L}{2} \label{eq:sub1} \\
L_{2} & \rightarrow  & n_{2} \frac{L}{2} \, .  \label{eq:sub2}
\end{eqnarray}
To conserve the total optical length during the transfer, we choose
\begin{eqnarray}
n_{1} & = & n_{0}+\eta \label{eq:n1} \\
n_{2} & = & n_{0}-\eta \, . \label{eq:n2}
\end{eqnarray} 
Here, $n_{0}$ is the unperturbed refractive index of the waveguides and $\eta$ is the change due to the modulation. Even though the substitutions (\ref{eq:sub1}) and (\ref{eq:sub2}) capture the main relations between the two models, there remain some technical differences which originate with the second boundary condition given below Eq.\ (\ref{Wavemode}). We find that in the coupled waveguide case the transcendental equation (\ref{transcendental1}) for the global wave numbers becomes
\begin{equation}
\tan(n_{2} k_{m} L/2)=\frac{n_{2}}{n_{1}} \ \frac{\tan(n_{1} k_{m} L/2)}{( \alpha k_{m}/n_{1}) \tan(n_{1} k_{m} L/2)-1} \ .
\label{transcendental3}
\end{equation}
Although the arguments of the tangent functions are exactly as given by the above substitutions, the factors involving the refractive indices $n_{1}$ and $n_{2}$ that appear outside of the tangents are not. However, we note that in practice the magnitude of the perturbation $\eta$ that is required to change the optical length  $\Delta L=L_{1}-L_{2} \rightarrow \eta L $ by a significant amount can be small. Indeed, according to Section \ref{sec:transferratio}, complete transfer of a mode is achieved when $ k \Delta L \approx \pi $, and if we take $\lambda= 780$ nm and $L=100 \, \mu$m we then find we require $\eta=0.004$. This should be compared with $n_{0}$ which by definition must be greater than unity. To a good approximation one can then replace Eq.\  (\ref{transcendental3}) by
\begin{equation}
\tan(n_{2} k_{m} L/2)= \ \frac{\tan(n_{1} k_{m} L/2)}{( \alpha k_{m}/n_{0}) \tan(n_{1} k_{m} L/2)-1} \ .
\label{transcendental4}
\end{equation}
Note that we have kept the full refractive index dependence in the arguments of the tangent functions where it is crucial. This transcendental equation is obtained from Eq.\ (\ref{transcendental1}) by precisely the substitutions (\ref{eq:sub1}) and (\ref{eq:sub2}), plus the replacement  $ \alpha k \rightarrow \alpha k /n_{0}$. This latter replacement can be understood by examining the transmission properties of a $\delta$-function mirror that lies between two dielectric media, one with refractive index $n_{1}$, and the other with refractive index $n_{2}$. We find that
\begin{equation}
t= \frac{2 n_{1}}{n_{1}+n_{2}-ik \alpha}
\end{equation}
so that 
\begin{equation}
T=\vert t \vert^2 = \frac{n_{1}^2}{\frac{(n_{1}+n_{2})^2}{4}+\frac{k^2 \alpha^2}{4}}\approx \frac{1}{1+\frac{k^2 \alpha^2}{4 n_{0}^2}} \, ,
\end{equation}
cf.\ Eq.\ (\ref{eq:deltatransmission}). In the last step we substituted the definitions (\ref{eq:n1}) and (\ref{eq:n2}) and assumed that $\eta \ll n_{0}$. Thus, when surrounded by a medium of refractive index $n_{0}$, the transmission of the $\delta$-function mirror is obtained from that in vacuum by putting $\alpha k \rightarrow \alpha k /n_{0}$, as above. This means that to maintain the same transmission in the two cases one should increase either $k$ or $\alpha$ by a factor of $n_{0}$.

With a little care most of formulae given in the main body of this paper can be modified for the waveguide case. For example, in the key Eqns (\ref{even approx2}), (\ref{odd approx2}), (\ref{eq:gamma}) and (\ref{Delta approx}), the speed of light $c$ in needs to be replaced by the speed $c/n_{0}$ in  the guide.

\section{The diabatic basis}
\label{sec:appendix3}

The solution of the transcendental equation (\ref{transcendental2}) allows us to construct global modes which exist throughout the double cavity. The global modes are the eigenvectors of the time-independent Maxwell wave equation in the double cavity and need to be recalculated for each value of the difference between the cavity lengths $\Delta L$. These modes give the so-called adiabatic basis as specified by Eqns (\ref{eq:E+=keven}) and (\ref{eq:E-=kodd}), and we denote this basis by $U_{e}(x)$ and $U_{o}(x)$.

However, as discussed in Section \ref{sec:LZ}, it is convenient to be able to express the double cavity problem in a basis which we refer to as the ``LZ diabatic basis'', or the ``diabatic basis'' for short. We \emph{define} the diabatic basis as the basis in which solutions to the time-independent Maxwell/Schr\"{o}dinger wave equation can be expressed in the form given by the LZ hamiltonian matrix (\ref{LZ-Hamiltonian}). In other words, the diabatic energies $\pm E(t)$ and eigenvectors are those found from the LZ hamiltonian when $\Delta=0$. We shall denote the diabatic basis vectors $ \phi_{L}(x)$ and $\phi_{R}(x)$.

For a $\delta$-function mirror the diabatic basis does not in general coincide with the entirely localized basis, as defined by the wave numbers $k_{n}= 2 \pi n /(L\pm \Delta L)$ found when $\alpha \rightarrow \infty$. This can easily be seen by comparing the LZ hamiltonian (\ref{LZ-Hamiltonian}),  for which the diabatic energies cross exactly halfway between the two adiabatic solutions, to the entirely localized solutions which cross on the lower adiabatic curve (see Figure \ref{fig:avoidedcrossing}).

Actually, we are in the slightly curious situation of already knowing the adiabatic basis and needing to find the diabatic basis. This is the converse of the usual situation. Indeed, the adiabatic solutions are in a sense better defined because they correspond to the basis which diagonalizes the hamiltonian and is therefore unique. Meanwhile, the diabatic basis is somewhat arbitrary because it corresponds to a particular ``undiagonalization'' of the hamiltonian. Nevertheless, this particular undiagonalization into the specific LZ form (\ref{LZ-Hamiltonian}) can easily be achieved by rotating the states correctly. 

We begin from the diagonalized LZ matrix (i.e.\ LZ hamiltonian in the adiabatic basis)
\begin{equation}
H_{\mathrm{ad}}=\left[\begin{array}{cc}
E_{+}(t) & 0\\
0 & E_{-}(t) \end{array}\right]
\label{eq:diagonalized}
\end{equation}
where $E_{\pm}=\pm\sqrt{E^2+\Delta^2}$ are the adiabatic energies (global wave numbers), where the plus sign corresponds to the solution which is even at $\Delta L=0$. We want to transform this matrix into the LZ matrix in the diabatic basis
\begin{equation}
H_{LZ}=\left[\begin{array}{cc}
E(t) & \Delta\\
\Delta & -E(t) \end{array}\right] \ .
\label{eq:undiagonalized}
\end{equation}

In order to go between (\ref{eq:diagonalized}) and (\ref{eq:undiagonalized}) we perform a similarity transformation. Let $H_{\mathrm{ad}}=S^{-1}H_{LZ}S$, where 
\begin{equation}
S=\left[\begin{array}{cc}
\cos\theta & \sin\theta\\
-\sin\theta & \cos\theta\end{array}\right] \ .
\label{Similarity Transformation}
\end{equation}
We find that the matrix elements of $S$ are given by
\begin{equation}
\cos\theta=\sqrt{\frac{\sqrt{E^{2}+\Delta^{2}}+E}{2\sqrt{E^{2}+\Delta^{2}}}}=\sqrt{\frac{E_{+}+\sqrt{E_{+}^{2}-\Delta^{2}}}{2E_{+}}}
\label{cos}
\end{equation}
and
\begin{equation}
\sin\theta=-\sqrt{\frac{\sqrt{E^{2}+\Delta^{2}}-E}{2\sqrt{E^{2}+\Delta^{2}}}}=-\sqrt{\frac{E_{+}-\sqrt{E_{+}^{2}-\Delta^{2}}}{2E_{+}}}
\label{sin}
\end{equation}
where the expressions after the second equality in both equations are written in terms of the adiabatic quantities obtained by solving the transcendental equation (\ref{transcendental2}). In particular, the value of $\Delta$ is found from $\Delta=E_{+}\vert_{\Delta L=0}$.

The relationship between the two sets of basis states is then
\begin{eqnarray}
\phi_{R}(x) & = & \cos\theta \ U_{e}(x) +\sin\theta \  U_{o}(x) \label{phi-L}
\\
\phi_{L}(x)  & = & -\sin\theta \ U_{e}(x) +\cos\theta \ U_{o}(x)
\label{phi-R}
\end{eqnarray}
and conversely
\begin{eqnarray}
U_{e}(x) & = & \cos\theta \ \phi_{R}(x) -\sin\theta \ \phi_{L}(x) \label{eq:Ueven}
\\
U_{o}(x) & = & \sin\theta \  \phi_{R}(x)  +\cos\theta \ \phi_{L}(x)  \ .
\label{eq:Uodd}
\end{eqnarray}
The orthogonality relations satisfied by $\phi_{L/R}$ can be found by substituting Eqns (\ref{eq:Ueven}) and (\ref{eq:Uodd}) into Eq.\ (\ref{normalization}). One finds
\begin{eqnarray}
\frac{1}{\varepsilon_{0}}\int_{-L_{1}}^{L_{2}}\varepsilon(x)\phi_{L/R}(x)\phi_{L/R}(x)dx & = & 1 \label{eq:normalizationphi1}\\
\frac{1}{\varepsilon_{0}}\int_{-L_{1}}^{L_{2}}\varepsilon(x)\phi_{L}(x)\phi_{R}(x)dx & = & 0 \label{eq:normalizationphi2} \ .
\end{eqnarray}

\section{Time variation of the diabatic modes}
\label{sec:appendix4}

\begin{figure}
\includegraphics[width=1.0\columnwidth]{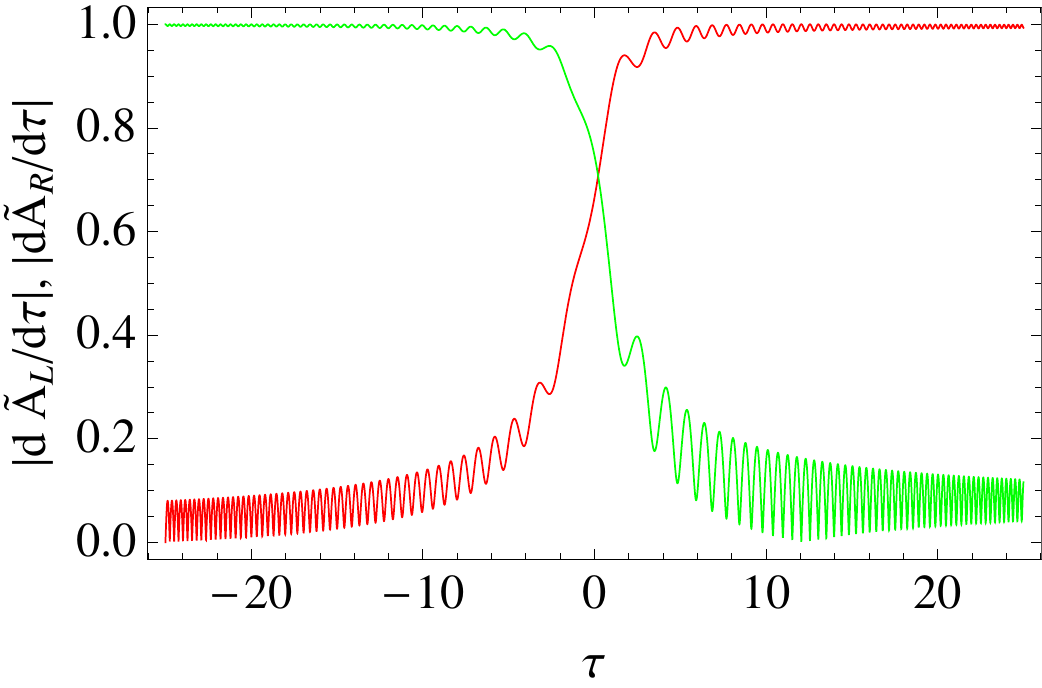}
\caption{(Color online) Time derivatives of the diabatic amplitudes, as obtained by numerical solution of Eq.\ (\ref{eq:schrod}). The green curve (that takes the value one at $\tau \equiv t \Delta / \hbar =-50$) gives the modulus of $\mathrm{d} \tilde{A}_{R}/\mathrm{d} \tau $, and the red curve gives the modulus of  $\mathrm{d} \tilde{A}_{L}/\mathrm{d} \tau $. The values of the parameters are set to $\tilde{\vartheta} \equiv \hbar \vartheta / \Delta^2=1$.}
\label{fig:LZtimederivative}
\end{figure}

In this Appendix we check the validity of the approximation made in Section \ref{sec:frommaxwelltoschrodinger} whereby  the time-dependence of the diabatic mode functions $\phi_{L}$ and $\phi_{R}$ is neglected in comparison to the time dependence of the diabatic amplitudes $A_{L}$ and $A_{R}$. If the time dependence of the diabatic modes is taken into account then the right hand side of Eq.\ (\ref{eq:maxwell2b}) becomes $\ddot{A}_{L}\phi_{L}+2\dot{A}_{L}\dot{\phi}_{L}+A_{L}\ddot{\phi}_{L}$ plus an identical term where $L \rightarrow R$. The largest terms we are neglecting in Section \ref{sec:frommaxwelltoschrodinger} are the cross terms $\dot{A}_{L/R}\dot{\phi}_{L/R}$. 

\begin{figure}
\includegraphics[width=1.0\columnwidth]{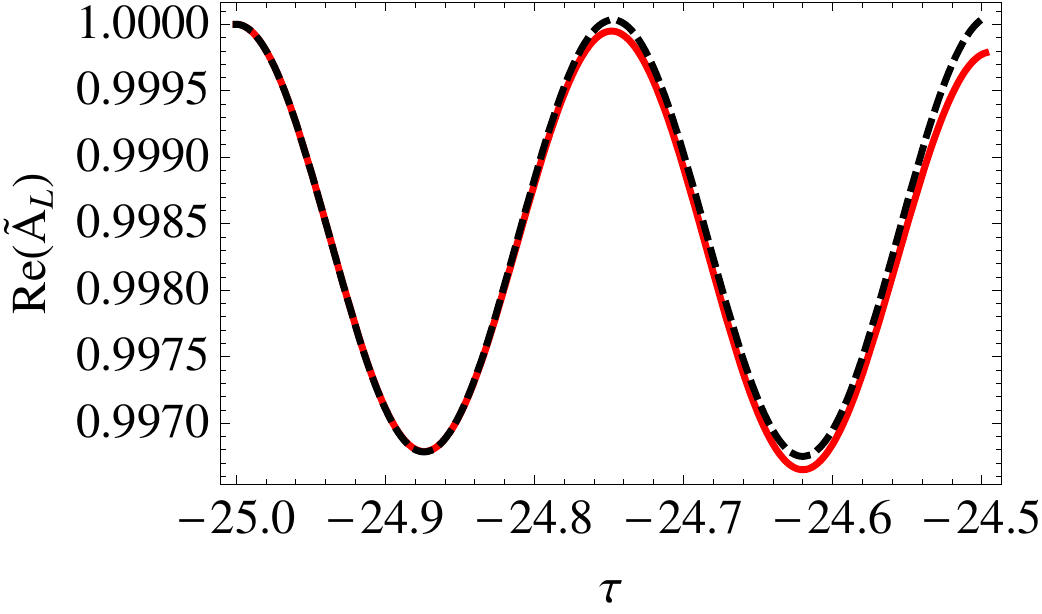}
\caption{(Color online) Early time behavior of the real part of $\tilde{A}_{L}$. The solid red curve gives the numerical solution of Eq.\ (\ref{eq:schrod}) and the dashed black curve gives the approximate analytic solution given by Eq.\ (\ref{eq:analyticsol}). The values of the parameters are the same as those in Fig.\ \ref{fig:LZtimederivative}.}
\label{fig:realsol}
\end{figure}

\begin{figure}
\includegraphics[width=1.0\columnwidth]{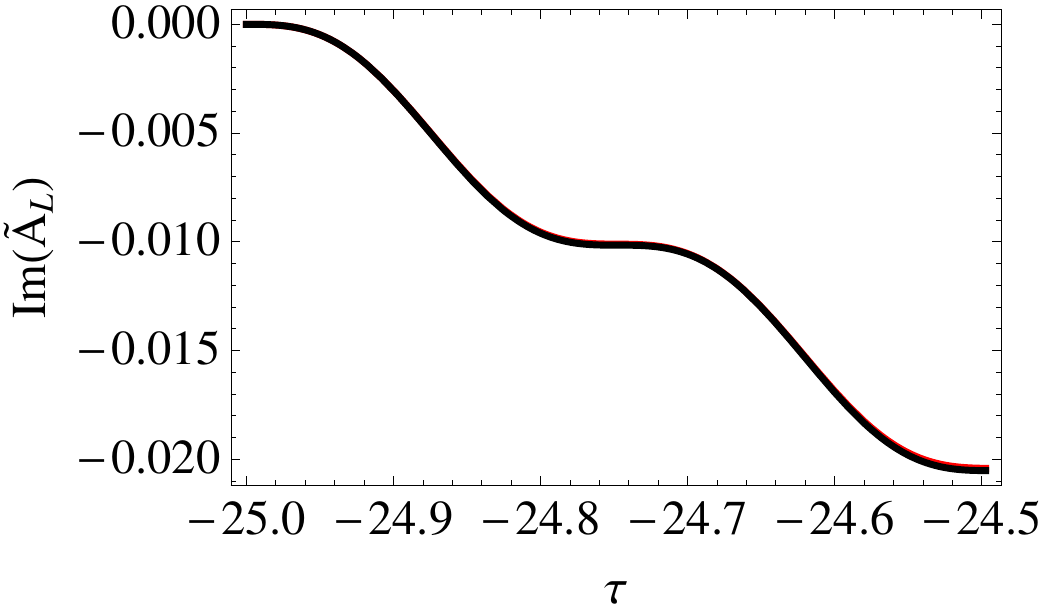}
\caption{(Color online) Early time behavior of the imaginary part of $\tilde{A}_{L}$. Two curves are plotted, one giving the numerical solution of Eq.\ (\ref{eq:schrod}) and the other giving the approximate analytic solution as given by Eq.\ (\ref{eq:analyticsol}), but the difference between the two is hardly visible for the times shown. Only for times greater than $\tau=-22$ do differences really become apparent.  The values of the parameters are the same as those in Fig.\ \ref{fig:LZtimederivative}.}
\label{fig:imagsol}
\end{figure}

Even for quite modest values of the reflectivity, the diabatic modes become strongly localized on one side of the mirror or the other allowing us to accurately approximate the diabatic mode functions by the completely localized mode functions given by
\begin{equation}
\phi_{L}(x) \approx \sqrt{\frac{4}{L+\Delta L}} \sin \left[ n \pi \left( \frac{2x}{L+\Delta L}+1\right) \right]
\end{equation}
and a similar expression for $\phi_{R}$. The cross terms referred to above can then be shown to lead to new terms in the equations of motion (i.e.\ once the mode functions have been integrated out) of the form 
\begin{equation}
\dot{A}_{L} \int_{-L_{1}}^{0} \dot{\phi}_{L} \phi_{L} \mathrm{d}x= \dot{A}_{L} v \int_{-L_{1}}^{0} \frac{\partial \phi_{L}}{\partial \Delta L}  \, \phi_{L} \, \mathrm{d}x \approx -\frac{v}{L} \dot{A}_{L}
\end{equation}
and a similar term for $\phi_{R}$. This needs to be compared with $\ddot{A}_{L/R}$. Removing the diagonal terms via the transformation (\ref{eq:Atilde}) to give the slowly rotating amplitudes, we find that, assuming $\Delta L \ll L$, the quantities that must be compared are
\begin{equation}
-2 i \beta_{L/R} \dot{\tilde{A}}_{L/R} \quad \mbox{vs} \quad 2(\dot{\tilde{A}}_{L/R}-i\beta_{L/R}\tilde{A}_{L/R})\frac{v}{L} 
\label{eq:comparison}
\end{equation}
where the left hand side refers to the existing calculation in the main part of this paper and the right hand side gives the new terms.
In the case when $v =  1$ m/s and $L=100 \, \mu$m, the first term on the right hand side is clearly negligible in comparison to the other terms. To find the magnitudes of the remaining terms let us go back to the Schr\"{o}dinger-like equations (\ref{eq:schrod}). Taking the derivative of these equations with respect to time we find, after some manipulation, the following two uncoupled equations for each amplitude 
\begin{eqnarray}
\frac{\mathrm{d}^2 \tilde{A}_{L}}{\mathrm{d} \tau^2} & = & -i \tilde{\vartheta} \tau \frac{\mathrm{d} \tilde{A}_{L}}{\mathrm{d} \tau}-\tilde{A}_{L} \label{eq:LZclosedL} \\
\frac{\mathrm{d}^2 \tilde{A}_{R}}{\mathrm{d} \tau^2} & = & i \tilde{\vartheta} \tau \frac{\mathrm{d} \tilde{A}_{R}}{\mathrm{d} \tau}-\tilde{A}_{R} \label{eq:LZclosedR}
\end{eqnarray}
where we have defined the dimensionless quantities $\tau \equiv t \Delta / \hbar$ and $\tilde{\vartheta} \equiv \hbar \vartheta/ \Delta^2$. In his original paper \cite{zener32}, Zener mapped these equations onto the Weber differential equation, and used the known asymptotic results for Weber functions for $\tau \rightarrow \pm \infty$ to obtain the transfer probability quoted in Eq.\ (\ref{eq:LZprobability}).
Here we are interested in a slightly different problem, namely the solutions of equations (\ref{eq:LZclosedL}) and (\ref{eq:LZclosedR}) for finite times. In particular, referring to Figure \ref{fig:LZtimederivative}, we see that the time derivative of $\tilde{A}_{L}$ takes its smallest values at the beginning of the time evolution. These are the times we therefore need to worry about most since it is then that it is most likely that the magnitude of the right hand side of Eq.\ (\ref{eq:comparison}) may exceed the magnitude of the left hand side.

At early times, when $A_{L} \approx 1$ and $A_{R} \approx 0$, we find that Equation (\ref{eq:LZclosedL}) is approximately satisifed by 
\begin{eqnarray}
\tilde{A}_{L}(\tau) & = & 1  +  \frac{i}{\tilde{\vartheta}} \ln[\tau/\tau_{0}] \label{eq:analyticsol} \\ && + \frac{i}{2 \tilde{\vartheta}}  \exp [i \tilde{\vartheta}\tau_{0}^2/2] \left(  \mathrm{Ei}[-i \tilde{\vartheta}\tau_{0}^2/2]   -  \mathrm{Ei}[-i \tilde{\vartheta} \tau^2/2] \right)  \nonumber
\end{eqnarray}
where $\mathrm{Ei}(z)$ is the exponential  integral function \cite{a+s}, and $\tau_{0}$ is the initial time at which $\tilde{A}_{L}=1$. The real and imaginary parts of this solution are compared to the numerical results in Figures (\ref{fig:realsol}) and (\ref{fig:imagsol}). This solution yields the following time derivative
\begin{equation}
\frac{\mathrm{d}\tilde{A}_{L}}{\mathrm{d} \tau}=\frac{i}{\tau \tilde{\vartheta}} \left(1-\exp[i \tilde{\vartheta} (\tau_{0}^{2}-\tau^2)/2] \right) \, .
\end{equation}

In terms of the dimensionless time $\tau$, the comparison given in Eq.\ (\ref{eq:comparison}) becomes (after dropping the negligible first term on the right hand side)
\begin{equation}
\frac{\mathrm{d}\tilde{A}_{L}}{\mathrm{d} \tau} \quad \mbox{vs} \quad \tilde{A}_{L}\frac{v}{L} \frac{\hbar}{\Delta} 
\label{eq:comparison2}
\end{equation}
From Eq.\ (\ref{eq:analyticsol}) we see that the magnitude of $\tilde{A}_{L}$ is approximately one, whereas the magnitude of  $\mathrm{d} \tilde{A}_{L} / \mathrm{d}\tau$ is approximately $1/\tau \tilde{\vartheta}$. Also, in the main part of the text we argued that a reasonable value of $\Delta/\hbar$  is $10^{11}\mathrm{s}^{-1}$. We therefore see that for the case considered in this paper (initial times $\tau_{0}=-25$ and $\tilde{\vartheta}=1$), the time dependence of the diabatic mode functions can be ignored.

\section{The Maxwell wave equation in a moving dielectric}
\label{sec:appendix5}
Leonhardt and Piwnicki \cite{leonhardt99} have derived a modified version of Maxwell's wave equation 
\begin{equation}
\left( \nabla^{2} -\frac{n_{r}^2}{c^2}\frac{\partial^{2}}{\partial t^2} -2 \frac{n_{r}^2-1}{c^2} \mathbf{v} . \nabla \frac{\partial}{\partial t} \right)\mathcal{E}(\mathbf{x},t)=0 \, ,
\label{eq:}
\end{equation}
in order to describe light propagating in a medium which is moving with velocity $\mathbf{v}$. This equation, which is valid in dispersionless dielectrics with refractive index $n_{r}(x,t)=\sqrt{\varepsilon(x,t)/\varepsilon_{0}}$, is  correct to first order in $\vert \mathbf{v} \vert /c$. In our case, the moving dielectric medium is the common mirror, and our aim here is to estimate the size of the last term on the left hand side because we have not included it in our treatment. The maximum size it can take is of order $k^2 v/c$, as can be seen by taking $2n_{r}(n_{r}^2-1)=1$ and letting both derivatives act on the phase $\phi=n_{r}kx-\omega t$ of the electric waves.  To justify our neglect of this term we must check that it is much smaller than the smallest term we keep in our reduction of Maxwell's wave equation to a first-order-in-time equation,  as described in Section \ref{sec:frommaxwelltoschrodinger}. Referring to Eq.\ (\ref{eq:ddotA}), we see that the smallest term we keep is $\beta \mathrm{d}{\tilde{A}}/\mathrm{d}t$, where $\beta \approx \omega_{\mathrm{av}}$. In Section \ref{sec:frommaxwelltoschrodinger} we argue that this term has a magnitude of $\beta \mathrm{d}{\tilde{A}}/\mathrm{d}t \approx  \omega_{\mathrm{av}} \Delta/\hbar = 10^{27} \,  \mathrm{s}^{-2}$, and thus $(\beta /c^2) \mathrm{d}{\tilde{A}}/\mathrm{d}t  \approx 10^{10} \, \mathrm{m}^{-2}$. This assertion  is supported by the results plotted in Fig.\ \ref{fig:LZtimederivative}, which show that $\vert \mathrm{d} \tilde{A}/ \mathrm{d} \tau \vert =  (\hbar/\Delta) \vert \mathrm{d} \tilde{A}/ \mathrm{d} t \vert \approx 1$. 

Consider first the case of a common mirror moved by piezoelectric motors, so that $v \approx 1$\, m/s. Then we have that $k^2  v/c \approx 10^6 \, \mathrm{m}^{-2}$, which is considerably smaller than the smallest retained term. However, the other case discussed in Section \ref{sec:experimentalfeasibility} concerned a waveguide whose refractive index is changed on a timescale of 10 ps, which is equivalent to a mirror moving at speeds $ v \approx 20,000$ m/s. Under these circumstances we  expect  the modifications to Maxwell's wave equation to become important.

\end{document}